\documentclass[12pt]{article}
\usepackage{geometry}
\usepackage{pdfsync}
\usepackage{graphicx}
\usepackage{amsmath}
\usepackage{amsfonts}
\usepackage{amssymb}
\usepackage{cite}
\usepackage{epstopdf}
\usepackage{comment}
\usepackage{cancel}
\usepackage{mathrsfs}
\usepackage{dsfont}

\usepackage{cleveref}
\usepackage{cite}
\crefname{section}{§}{§§}
\Crefname{section}{§}{§§}
\newcommand{\mfrac}[2]{{#1/#2}}
\numberwithin{equation}{section}
\usepackage[titletoc,title]{appendix}
\usepackage{cleveref}

\usepackage{booktabs} 
\newcommand{\bea}{\begin{eqnarray}}

\newcommand{\eea}{\end{eqnarray}}

\newcommand{\be}{\begin{equation}}
\newcommand{\ee}{\end{equation}}
\newcommand{\ba}{\begin{align}}
\newcommand{\ea}{\end{align}}

\newcommand{\nn}[0]{\nonumber}

   \makeatletter
  \let\over=\@@over \let\overwithdelims=\@@overwithdelims
  \let\atop=\@@atop \let\atopwithdelims=\@@atopwithdelims
  \let\above=\@@above \let\abovewithdelims=\@@abovewithdelims
\renewcommand\section{\@startsection {section}{1}{\z@}%
                                   {-3.5ex \@plus -1ex \@minus -.2ex}%nn
                                   {2.3ex \@plus.2ex}%
                                   {\normalfont\large\bfseries}}

\renewcommand\subsection{\@startsection{subsection}{2}{\z@}%
                                     {-3.25ex\@plus -1ex \@minus -.2ex}%
                                     {1.5ex \@plus .2ex}%
                                     {\normalfont\bfseries}}

\newcommand{\ophi}[0]{\overline{\phi}}
\newcommand{\oPhi}[0]{\overline{\Phi}}
\newcommand{\oF}[0]{\overline{F}}
\newcommand{\of}[0]{\overline{f}}
\newcommand{\oh}[0]{\overline{h}}
\newcommand{\oH}[0]{\overline{H}}
\newcommand{\oeta}[0]{\overline{\eta}}
\newcommand{\oell}[0]{\overline{\ell}}
\newcommand{\oz}[0]{\overline{z}}
\newcommand{\oy}[0]{\overline{y}}
\def\mystrut{\vrule height 12pt depth 1pt width 0pt}

\def\simlt{\mathrel{\lower2.5pt\vbox{\lineskip=0pt\baselineskip=0pt
           \hbox{$<$}\hbox{$\sim$}}}}
\def\simgt{\mathrel{\lower2.5pt\vbox{\lineskip=0pt\baselineskip=0pt
           \hbox{$>$}\hbox{$\sim$}}}}

\allowdisplaybreaks
\begin{document}

\setcounter{page}{1}

\begin{titlepage}
\begin{flushright}
ACT-05-21\\ MI-HET-768
\end{flushright}

\unitlength = 1mm~%\\
%\vskip .2cm
\begin{center}

{\LARGE{\textsc %Aspects 
Particle physics and cosmology of the string derived no-scale 
flipped $SU(5)$ %particle cosmology
}}

\vspace{0.3cm}
 {\large I. Antoniadis}\,{}\footnote{{\tt antoniad@lpthe.jussieu.fr}}$^{a}$  {\large D.V. Nanopoulos}\,{}\footnote{\tt dimitri@physics.tamu.edu}$^{b}$   {\large J. Rizos}\,{}\footnote{\tt irizos@uoi.gr}${}^c$

\vspace{0.2cm}

{\it  ${}^a$ Laboratoire de Physique Th\'eorique et Hautes Energies - LPTHE\\ Sorbonne Universit\'e, CNRS, 4 Place Jussieu, 75005 Paris, France; \\ 
Department of Mathematical Sciences, University of Liverpool,\\ Liverpool L69 7ZL, United Kingdom\\
	  ${}^b$ George P. and Cynthia W. Mitchell Institute for Fundamental Physics and Astronomy, Texas A\&M University, College Station, TX 77843, USA; \\
Astroparticle Physics Group, Houston Advanced Research Center (HARC)\\ Mitchell Campus, Woodlands, TX 77381, USA;\\
Academy of Athens, Division of Natural Sciences, Athens 10679, Greece\\
${}^c$ Physics Department, University of Ioannina, 45110, Ioannina, Greece;\\
School of Science and Technology, Hellenic Open University\\
Tsamadou 13-15, GR-26222 Patras, Greece 
\\
}

\vspace{0.2cm}

\begin{abstract}
In a recent paper, we identified a cosmological sector of a flipped $SU(5)$ model derived in the free fermionic formulation of the heterotic superstring, containing the inflaton and the goldstino superfields with a superpotential leading to Starobinsky type inflation, while $SU(5){\times}U(1)$ is still unbroken.
Here, we study the properties and phenomenology of the vacuum after the end of 
inflation, where the gauge group is broken to the Standard Model. We identify a 
set of vacuum expectation values, triggered by the breaking of an anomalous 
$U(1)_A$ gauge symmetry at roughly an order of magnitude below the string 
scale, that solve the F and D-flatness supersymmetric conditions up to 6th 
order in the superpotential which is explicitly computed, leading to a 
successful particle phenomenology. In particular, all extra colour triplets 
become superheavy guaranteeing observable proton stability, while the Higgs 
doublet mass matrix has a massless pair eigenstate with realistic hierarchical 
Yukawa couplings to quarks and leptons. The supersymmetry breaking scale is 
constrained to be high, consistent with the non observation of supersymmetric 
signals at the LHC. 

\end{abstract}

\setcounter{footnote}{0}
\vspace{.2cm}
\end{center}

\end{titlepage}

\pagestyle{empty}
\pagestyle{plain}

\pagenumbering{arabic}

\tableofcontents
\bibliographystyle{utphys}

\newpage
\section{Introduction}\label{intro}
In this work, we make an important step further on the avenue we started recently~\cite{Antoniadis:2020txn}, towards a string derived microscopic model that provides a simultaneous description of fundamental particle physics and cosmology. The model was constructed in 1989~\cite{Antoniadis:1989zy} within the framework of free fermionic formulation of four-dimensional (4d) heterotic superstring~\cite{Antoniadis:1986rn} and has an observable sector based on the flipped $SU(5)\times U(1)$ gauge group with three chiral families of quarks and leptons \cite{Barr:1981qv,Antoniadis:1987dx}. The basis vectors of boundary conditions defining the model, as well as its full massless spectrum are given again for self consistency in Appendix~A.

In~\cite{Antoniadis:2020txn}, we identified the inflaton among the gauge 
singlet massless states of the model with the superparner of a fermion mixed 
with the Right-handed neutrinos~\cite{Ellis:2013nxa}. It acquires a 
superpotential together with the goldstino at 6th and 8th order in the string 
slope $\alpha'$-expansion, via vacuum expectation values (VEVs) of fields 
generated by the breaking of an anomalous $U(1)_A$ gauge symmetry 
(characteristic in heterotic models~~\cite{Dine:1987xk}) at a calculable scale 
an order of magnitude below the string mass, related to the anomaly. This scale 
in string units introduces therefore a small parameter allowing perturbative 
computations around the free-fermionic point where all fields and moduli are 
fixed at zero VEVs. As a result, the inflation scale turns out to be about five 
orders of magnitude lower, in the range of $10^{13}$ GeV, while the 
superpotential leads to a Starobinsky-type inflation~\cite{starobinsky} due to 
the no-scale structure of the low energy effective 
supergravity~\cite{Cremmer:1983bf} which is calculable in our model to all 
orders in $\alpha'$~\cite{Antoniadis:2020txn, Antoniadis:1987zk, 
Ferrara:1987tp}. Note that during inflation, $SU(5)\times U(1)$ remains 
unbroken because its breaking occurs via a first order phase transition at a 
critical temperature which is lower that the scale of 
inflation~\cite{Ellis:2018moe, Ellis:2019opr, Ellis:2020lnc}.

In this work, we extend the previous analysis to the study of the vacuum of the theory, after the end of inflation, where $SU(5)\times U(1)$ is broken to the Standard Model. More precisely, we find a consistent set of VEVs that solve the F and D-flatness equations up to 6th order in the $\alpha'$-expansion of the superpotential. This vacuum obviously preserves supersymmetry, whose breaking we don't discuss here. An important result to emphasise is that the requirement of gauge symmetry breaking by a pair of ${\bf 10}+{\bf\overline{10}}$ leads to a slight reorganisation of the choice of VEVs for the $SU(5)\times U(1)$ gauge singlet states, compared to the set we had during the inflationary phase, in order to satisfy the flatness conditions.

Our analysis for finding the choice of VEVs requires in some cases the knowledge of the exact coefficients of higher dimensional operators, or precise relations among them. Thus, besides applying selection rules to find which of those are non-vanishing, we need to perform some explicit computations of superpotential terms to 5th or 6th order which is highly non-trivial. The challenging part involves correlation functions of several primary operators in the Ising model for which we concentrate a dedicated section of this paper.

Obviously, one of our goals is to identify the quarks and leptons among the three chiral generations and one vector-like pair, as well as a pair of Higgs doublets with the required Yukawa couplings. On the other hand, one should also ensure that all colour triplet states acquire masses at high scale so that there is no dangerous proton decay. For these reasons, we impose in our choice of VEVs that the colour triplet mass matrix has a non-vanishing determinant, while the weak doublet matrix has exactly one massless eigenstate with components along the doublets that provide a realistic hierarchical Yukawa matrix, taking into account the successful phenomenological analysis of the model that has been done in the past~\cite{Antoniadis:1989zy, Lopez:1989fb, Lopez:1991ac, Rizos:1990xn, Ellis:1990vy, Antoniadis:1991fc}.
Moreover, the constraints from proton decay impose that the supersymmetry breaking scale $m_{\rm susy}$ should be at least of order of tens of TeV, compatible with an independent analysis of reheating and nucleosynthesis requiring $m_{\rm susy}$ to lye in this energy region~\cite{Ellis:2019opr}, possibly within the reach of the next generation of high energy hadron colliders.

The outline of the paper is the following. In Section~\ref{scosmology}, we review briefly our previous results~\cite{Antoniadis:2020txn}, such as the identification of the inflaton and the goldstino, as well as the choice of VEVs giving rise to the inflationary superpotential. In Section~\ref{doublet-triplet}, we impose the phenomenological constraints realising the triplet-double splitting at the string level, making all colour triplets superheavy while leaving massless one pair of massless doublets. In Section~\ref{FDsol}, we compute and solve the D and F-flatness conditions up the 6th order, taking into account the above constraints. Section~5 is devoted to the explicit computation of some superpotential coefficients involving higher point functions of primary operators in the Ising model. In Section~6, we perform the phenomenological analysis by identifying the electroweak Higgs doublets and the quarks and leptons, and by computing in particular the structure of fermion masses. We also study proton decay by computing the relevant dimension-five operators induced by the Higgs triplets exchange, as well as those emerging directly at the string level. Finally, Section~7 contains some concluding remarks. The paper has also three Appendices. Appendix~A contains a brief summary of the `revamped' flipped $SU(5)$ string model,  Appendix~B contains the list and details on the flatness conditions up to 5th order in the superpotential which we use in our analysis in Section~\ref{FDsol}, while Appendix~C contains operator product expansions and various correlators of the Ising model that we use in Section~\ref{NRTcomputation}.

\section{The revamped flipped model and its cosmology sector}
\label{scosmology}
For convenience of the reader, we recall briefly the massless spectrum of the `revamped' flipped $SU(5)$ model (with the original notation of~\cite{Antoniadis:1989zy}), which consist of:
\begin{enumerate}
\item A gauge group $[SU(5)\times U(1)]\times U(1)^4\times[SO(10)\times SO(6)]$, where besides the observable and hidden sectors in the first and last bracket, there are four $U(1)$s, one combination of which is anomalous and becomes massive.
\item Four generations and one anti-generation of chiral matter in the representations ${\bf 10}+{\bf\bar 5}+{\bf 1}$ (and ${\bf\overline{10}}+{\bf 5}+{\bf\bar 1}$) of $SU(5)\times U(1)$, denoted as $F_i,\bar{f}_i,l_i^c$ (and $\oF_5, f_4, \bar{\ell}^c_4$) where $i$ labels the corresponding vector of boundary conditions.
\item Four pairs of ${\bf 5}+{\bf\bar 5}$ containing the electroweak Higgs doublets, denoted by $(h_i,\bar{h}_i)_{i=1,2,3}$ for those coming from the Neveu-Schwarz (NS) sector and $(h_{45}+\bar{h}_{45})$ for the pair coming from the $b_4+b_5$ sector.
\item Five vectors $({\bf 10},{\bf 1})+({\bf 1},{\bf 6})$ of the hidden gauge group, denoted as $T_i$ and $D_i$, respectively.
\item Six pairs $({\bf 4}+{\bf\bar{4}})$ of $SO(6)$ with fractional electric charges $\pm 1/2$, denoted as $(Z_1+\bar{Z}_1)$, $(Z_2+\bar{Z}_2)$, $(Y_1+\bar{Y}_1)$, $(X_1+\bar{X}_1)$, $Y_2,Y_2'$, $\bar{X}_2,\bar{X}_2'$.
\item Ten pairs of non-abelian gauge singlets but charged under $U(1)^4$, denoted by $\Phi_{12},\Phi_{23},\Phi_{31}$ (with their conjugates) for those coming from the NS sector, and $\phi_1,\dots,\phi_4,\phi_{\pm},\phi_{45}$ (with their conjugates) for those coming from the $b_4+b_5$ sector.
\item Five gauge singlets from the NS sector, denoted as $\Phi_1,\dots,\Phi_5$.
\end{enumerate} 

The full tree-level (trilinear) superpotential reads
\begin{align}
W_3 =\, %&= \nonumber\\
& g_s \sqrt{2}\left[\mystrut F_1 F_1 h_1 + F_2 F_2 h_2 + F_4 F_4 h_1 + \oF_5 \oF_5 \oh_2 
+F_4 \of_5 \oh_{45}+F_3 \of_3 \oh_3\right.\nonumber\\
&
+\of_1 \ell^c_1 h_1+\of_2 \ell^c_2 h_2 +\of_5 \ell^c_5 h_2  + f_4 \oell^c_4 \oh_1\nonumber\\
& +
\frac{1}{\sqrt{2}} F_4 \oF_5 \phi_3 + \frac{1}{\sqrt{2}} f_4 \of_5 \ophi_2 + \frac{1}{\sqrt{2}} \oell^c_4 \ell^c_5 \ophi_2 
\nonumber\\
& + h_1 \oh_2 \Phi_{12} + \oh_1 h_2 \oPhi_{12} + h_2\oh_3\Phi_{23} + \oh_2 h_3 \oPhi_{23}
+ h_3 \oh_1\Phi_{31} \nonumber\\ 
&+ \oh_3 h_1\oPhi_{31}+h_3 \oh_{45}\ophi_{45} + \oh_3 h_{45} \phi_{45}
+ \frac{1}{2} h_{45} \oh_{45} \Phi_3 \nonumber\\
&+\phi_1\ophi_2 \Phi_4 + \ophi_1\phi_2 \Phi_4+  \phi_3 \ophi_4 \Phi_5 + \ophi_3 \phi_4 \Phi_5
+\frac{1}{2}\phi_{45} \ophi_{45} \Phi_3\nonumber\\
&+\frac{1}{2}\phi_{+} \ophi_{+} \Phi_3 
+\frac{1}{2}\phi_{-} \ophi_{-} \Phi_3 \label{wtree}\\
&+\Phi_{12} \Phi_{23} \Phi_{31} +\oPhi_{12} \oPhi_{23} \oPhi_{31}  +
\Phi_{12} \phi_+ \phi_- + \oPhi_{12} \ophi_+ \ophi_- 
\nonumber\\
&+
\frac{1}{2} \Phi_3 \sum_{i=1}^4 \phi_i \ophi_i +  \Phi_{12} \sum_{i=1}^4 \phi_i^2 
+  \oPhi_{12} \sum_{i=1}^4 \ophi_i^2\nonumber\\
&+D_1^2\oPhi_{23}+D_2^2\Phi_{31}+D_4^2\oPhi_{23}+D_5^2\oPhi_{31}+\frac{1}{\sqrt{2}} D_4 D_5 \ophi_3 \nonumber\\
&+T_1^2\oPhi_{23}+T_2^2\Phi_{31}+T_4^2\Phi_{23}+T_5^2\Phi_{31}+\frac{1}{\sqrt{2}} T_4 T_5 \phi_2 \nonumber\\
&+\frac{1}{\sqrt{2}} Y_1 \overline{X}_2 \phi_4 +\frac{1}{\sqrt{2}} Y_2 \overline{X}_1 \phi_1 + Y_2 \overline{X}_2\phi_+ 
+ \frac{1}{2} Z_1 \overline{Z}_1 \Phi_3\nonumber\\
&\left.+Z_2 \overline{Z}_2 \oPhi_{12}+Z_1 \overline{X}_2' \ell^c_2 + Y_2' Z_1 D_1\mystrut\right]\,,\nonumber
\end{align}
where $g_s$ is the string coupling.

We now summarise the main results we obtained on the cosmology sector of the model~\cite{Antoniadis:2020txn}. The identification of the inflaton comes from the single superpotential coupling of the form $\bf{10}$ $\bf{\overline{10}}$ $\bf{1}$ with the singlet being a R-handed neutrino, which is $F_4\bar{F}_5\phi_3$, implying that $\phi_3$ should be component of a linear combination of fields defining the inflaton $y$. Since $\phi_3$ corresponds to a state from the third twisted sector, inspection of the derived no-scale K\"ahler potential that leads to the required Starobinsky type inflation implies that the goldstino superfield $z$ should come from the third untwisted NS sector. Moreover, the inflationary superpotential should be (in supergravity units) 
\be
W_I=M_I z(y-\lambda y^2)\,,
\label{Winfl}
\ee
with $M_I$ the scale of inflation and $\lambda$ a parameter close to 1\footnote{The value $\lambda=1$ corresponds to the scalaron with the properties of an $R^2$ term in the effective action.}. Following a detailed analysis of the various possibilities and a study of the relevant non-renormalisable (NR) superpotential interactions up to 10th order in the $\alpha'$-expansion, we identified the goldstino with the gauge singlet superfield $\Phi_4$ from the NS sector, while the inflaton is a linear combination of $\phi_3$ and $\bar\phi_3$:
\begin{align}
y &=\sin\omega\,\phi_3-\cos\omega\,\bar{\phi}_3\quad;\quad 
\tan\omega=\langle\phi_4\rangle/\langle\bar\phi_4\rangle\label{y-id}\,,\\ 
z &=\Phi_4\,. \label{z-id}
\end{align}
The orthogonal linear combination to $y$ is massive, while phenomenological constraints on appropriate reheating through the inflaton decay into neutrinos, baryogenesis and light  neutrino masses require $\langle\phi_4\rangle/\langle\bar\phi_4\rangle\sim 10^{-3}$~\cite{Ellis:2018moe, Ellis:2019opr, Ellis:2020lnc}. 

The superpotential \eqref{Winfl} is generated at the 6th and 8th order and reads:
\begin{align}\label{WI}
W_I &= g_s %\sqrt{2} 
C_6 \left(g_s\sqrt{2\alpha'}\right)^{3}\,\ophi_3 \Phi_4 \langle{D_1}\rangle{\cdot}\langle{D_5}\rangle\, \langle{T_1}\rangle{\cdot}\langle{T_4}\rangle\\
&\hspace{0.8cm}+g_s %\sqrt{2} 
C_8 \left(g_s\sqrt{2\alpha'}\right)^{5}\,\ophi_3^2\Phi_4\langle{D_1}\rangle{\cdot}\langle{D_4}\rangle\, \langle{T_1}\rangle{\cdot}\langle{T_4}\rangle\langle{\Phi_{31}}\rangle\,,\nonumber
%\\ &= \zeta^4 \Phi_4\left(\frac{\gamma}{g\sqrt{2\alpha'}}\phi_0 +  \delta \zeta \phi_0^2 \right)\,,
\end{align}
where $C_6$ and $C_8$ stand for the numerical values of the correlators associated to the $N=6$ and $N=8$ NR couplings, respectively. Note the presence of several hidden sector fields. The reason is that the trilinear superpotential involving only gauge singlets under the non-abelian gauge group is exact and does not receive $\alpha'$ corrections~\cite{Lopez:1990wt}.
It turns out that a typical VEV %$\langle\varphi\rangle$ 
satisfying the D-term conditions is of order $\xi$:
\be
%\langle\varphi\rangle 
\xi\sim {M_s\over 2\pi}\quad;\quad M_s\equiv{1\over\sqrt{2\alpha'}}\,.
\label{typicalVEV}
\ee
Thus, defining $\alpha_s\equiv g_s/2\pi$, one gets the inflation scale $M_I\simeq C_6\alpha_s^5M_s\sim 10^{13}$ GeV and $\lambda$ an order one tuneable parameter. The choice of VEVs that solve the F and D-flatness conditions, giving rise to the above superpotential are given in the left panel of Table~\ref{tableofvevs}.

\begin{table}
\begin{tabular}{|c|c|c|}
\hline
Field&Assignments in Ref. \cite{Antoniadis:2020txn}&Assignments in Section 4\\
\hline
$\Phi_{12}$&0&0\\
\hline
$\oPhi_{12}$&0&$\xi^3$\\
\hline
$\Phi_{23}$&0&$\xi^3$\\
\hline
$\oPhi_{23}$&0&$\xi^3$\\
\hline
$\Phi_{31}$&$\xi$&$\xi$\\
\hline
$\oPhi_{31}$&0&$\xi$\\
\hline
$\Phi_{I}$&$0$&$0$\\
\hline
$\phi_{45}$&$\xi$&$\xi$\\
\hline
$\ophi_{45}$&$\xi^3$&$\xi^3$\\
\hline
$\phi_1$&$0$&$0$\\
\hline
$\phi_2$&$0$&$0$\\
\hline
$\phi_3$&$0$&$0$\\
\hline
$\phi_4$&$\xi^4$&$\xi$\\
\hline
$\ophi_1$&$0$&$\xi$\\
\hline
$\ophi_2$&$0$&$0$\\
\hline
$\ophi_3$&$0$&$0$\\
\hline
$\ophi_4$&$\xi^2$&$\xi^2$\\
\hline
$\phi_+$&$\xi$&$\xi$\\
\hline
$\phi_-$&$\xi^3$&$\xi$\\
\hline
$\ophi_+$&$\xi^3$&$\xi$\\
\hline
$\ophi_-$&$\xi$&$\xi$\\
\hline
$F_1$&$0$&$\xi^{3/2}$\\
\hline
$\oF_5$&$0$&$\xi^{3/2}$\\
\hline
$F_3$&$0$&$\xi^3$\\
\hline
$D_1$&$\xi$&$\xi$\\
\hline
$D_2$&$0$&$0$\\
\hline
$D_3$&0&$\xi^2$\\
\hline
$D_4$&$\xi$&$\xi$\\
\hline
$D_5$&$\xi$&$0$\\
\hline
$T_1$&$\xi$&$\xi$\\
\hline
$T_2$&$0$&$0$\\
\hline
$T_3$&$\xi$&$\xi^2$\\
\hline
$T_4$&$\xi$&$\xi^2$\\
\hline
$T_5$&$0$&$0$\\
\hline
&$D_1^2=D_4^2=D_5^2=0\ D_4{\cdot}D_5\sim\xi^4$&$D_1^2+D_4^2+T_1^2=0$\\
&$T_1^2=T_3^2=T_4^2=T_3{\cdot}T_4=0$&$T_4^2=T_1{\cdot}T_4=0$\\
\hline
\end{tabular}
\caption{\label{VEVs} Field VEV assignments in the F/D-flatness solution of Ref. \cite{Antoniadis:2020txn} (column 2)  and  the solution derived here, in Section~\ref{FDsol} (column 3). }
\label{tableofvevs}
\end{table}

\section{Triplet-doublet splitting in $SU(5)\times U(1)$ string model}
\label{doublet-triplet}
In many grand unified theories (GUTs),  including $SU(5)$ and $SO(10)$,  
Standard Model (SM) Higgs doublets reside in the lowest dimensional gauge group 
representation(s) together with colour triplets.
Nevertheless, a successful phenomenological model has to differentiate between Higgs doublets and additional triplets.  The former have to stay light down to the electroweak scale in order to realise the electroweak symmetry breaking and provide fermion masses. The latter mediate nucleon decay and, unless sufficiently heavy, they could lead to proton lifetimes incompatible with existing data. 
This is often referred to as the doublet-triplet splitting problem. 
Among possible solutions one singles out 
 the so-called missing partner mechanism that involves the breaking of the GUT gauge symmetry
via Higgs fields that include unequal numbers of SM colour triplets and electroweak doublets. 
The missing partner scenario is elegantly realised  in the  flipped $SU(5)$ model \cite{Antoniadis:1987dx}.
In the minimal case, the GUT breaking Higgs fields entail a pair of $\mathbf{10},\overline{\mathbf{10}}$ representations that include one pair of d-type colour triplets $\mathbf{3},\overline{\mathbf{3}}$ but no SM Higgs doublets. Interaction terms of these Higgs fields with a single pair of $\mathbf{5},\overline{\mathbf{5}}$ representations,
 $\mathbf{10}\times\mathbf{10}\times\mathbf{5}+ \overline{\mathbf{10}}\times\overline{\mathbf{10}}\times \overline{\mathbf{5}}$, provide masses for the triplet components leaving the associated doublets in the  $\mathbf{5},\overline{\mathbf{5}}$ unaffected, see e.g. \cite{Ellis:2020qad}.

A more complicated situation arises in the string implementation of the flipped $SU(5)$ model \cite{Antoniadis:1989zy}. Here, we have four pairs of $\mathbf{5},\overline{\mathbf{5}}$ 
fields, namely $h_i,\overline{h}_i, i=1,2,3,45$ and one pair of GUT breaking $\mathbf{10},\overline{\mathbf{10}}$  Higgs fields, referred as $F, \overline{F}$, where $F$ stands for a linear combination of $F_\alpha, \alpha=1,2,3,4$ and $\bar{F}\equiv\bar{F}_5$. SM Higgs doublets and additional triplets are assigned to $SU(5)\times{U(1)}$ representations as follows
\begin{align}
h_i\left(\mathbf{5},-1\right)&= H_i\left(\mathbf{1},\mathbf{2},-\frac{1}{2}\right)+ D_i\left(\mathbf{3},\mathbf{1},-\frac{1}{3}\right)\,,\\
F\left(\mathbf{10},+\frac{1}{2}\right)&= Q_H\left(\mathbf{3},\mathbf{2},+\frac{1}{6}\right)+ d^c_H\left(\overline{\mathbf{3}},\mathbf{1},+\frac{1}{3}\right)+ \nu^c_H\left(\mathbf{1},\mathbf{1},0\right)\,,
\end{align}
and similarly for the conjugate fields $\overline{h}_i$, $\overline{F}$. Taking into account the tree-level superpotential \eqref{wtree}, the Higgs doublet mass matrix reads
\begin{align}
	M_H^{(3)} = \bordermatrix{
		~&H_1&H_2&H_3&H_{45}\cr
		\bar{H}_1&0&\oPhi_{12}&\Phi_{31}&0\cr
		\bar{H}_2&\Phi_{12}&0&\oPhi_{23}&0\cr
		\bar{H}_3&\oPhi_{31}&\Phi_{23}&0&\phi_{45}\cr
		\bar{H}_{45}&0&0&\ophi_{45}&\Phi_3
	}\,,
	\label{deth3}	
\end{align}
where $\Phi_3$, $\Phi_{12},\oPhi_{12}, \Phi_{23}, \oPhi_{23}, \Phi_{31}, \oPhi_{31}$  stand for flipped $SU(5)$ singlets that can acquire VEVs.

The associated triplet mass matrix depends on the choice of $F$, which is the linear combination of the $F_\alpha$ that acquires VEV. 
Notice, that there are 
two candidate up-quark couplings at the tree-level superpotential \eqref{wtree}, namely $F_4 \of_5 \oh_{45}$ and  $F_3 \of_3 \oh_{3}$.
Taking into account the absence of associated down-quark couplings of the form 
$F_3 F_3 h_i$, $i=1,2,3,45$, both at tree-level and at higher order $N=4,5$ 
superpotential terms~\cite{Lopez:1989fb, Rizos:1990xn, Ellis:1999ce}, we are 
led to identify the heaviest generation of quarks and leptons with $F_4, 
\of_5$. Moreover, in \eqref{wtree} there exist three terms providing mass to 
the down quarks  namely $F_1 F_1 h_1$, $F_4 F_4 h_1$ and $F_2 F_2 h_2$. Bearing 
in mind that the first two involve the same doublet $H_1$ (residing in $h_1$) 
and $F_4$ is assigned to the third generation we deduce that $F_1$ has to be 
associated with the GUT breaking Higgs field $F$.
In this case the additional triplet mass matrix can be recast in the form
\begin{align}
	M_D^{(3)} = \bordermatrix{
		~&D_1&D_2&D_3&D_{45}&\overline{d}^c_H\cr
		\bar{D}_1&0&\oPhi_{12}&\Phi_{31}&0&0\cr
		\bar{D}_2&\Phi_{12}&0&\oPhi_{23}&0&2\oF_5\cr
		\bar{D}_3&\oPhi_{31}&\Phi_{23}&0&\phi_{45}&0\cr
		\bar{D}_{45}&0&0&\ophi_{45}&\Phi_3&0\cr
		d^c_H&2F_1&0&0&0&0\cr
	}\,,
		\label{tripmm}
\end{align}
where $d^c_H, \overline{d}^c_H$ are the triplets residing in $F_1, \oF_5$ respectively. 

The doublet-triplet splitting conditions can now be expressed as
\begin{align}
\det{M_H^{(3)}} = 0\ ,\  \det{M_D^{(3)}} \ne 0\,,
\label{dtcon}
\end{align}
where
\begin{align}
\det{M_H^{(3)}}&=\left(\Phi_{12} \Phi_{23} \Phi_{31}+\oPhi_{12} \oPhi_{23} \oPhi_{31}\right)\Phi_3 + 
\Phi_{12} \oPhi_{12} \phi_{45} \ophi_{45} 
\,,
\\
\det{M_D^{(3)}}&=-4 F_1\oF_5\left(
\Phi_{23}\Phi_{31}\Phi_3
+\phi_{45}\ophi_{45}\oPhi_{12}
\right)\,,
\label{detmd}
\end{align}
are the determinants of relevant mass matrices. Additional constraints 
arise from low energy phenomenology and more particularly from the requirement to have a top quark mass coupling at the tree-level superpotential and in particular from the aforementioned coupling $F_4 \of_5 \oh_{45}$. Expressing the massless Higgs doublet 
pair as
\begin{align}
H &= c_1 H_1 + c_2 H_2 + c_3 H_3 + c_{45} H_{45}\,,\\
\bar{H} &= \bar{c}_1 \bar{H}_1 + \bar{c}_2 \bar{H}_2 + \bar{c}_3 \bar{H}_3 + \bar{c}_{45} \bar{H}_{45}\,,
\end{align}
%with $\sum_{i} c_i^2=\sum_i\bar{c}_i^2 =1$.
where $c_i, \overline{c}_{i}$ depend on the $SU(5)\times{U(1)}$ singlet VEVs, this requirement is translated to
\begin{align}
\bar{c}_{45}\ne0\label{tcon}
\,.
\end{align}

The constraints \eqref{dtcon}, \eqref{tcon} depend on $F_1, \oF_5$ and nine additional parameters, namely $\Phi_{12}, \oPhi_{12}, \Phi_{23},\oPhi_{23},\Phi_{31},\oPhi_{31},\phi_{45},\ophi_{45},\Phi_3\,$.
As these VEVs are subject to non-trivial F/D flatness conditions
we seek for solutions of \eqref{dtcon}, \eqref{tcon} with the minimum number of vanishing VEVs and no fine-tunings.
After some algebra we find no solution with a single vanishing VEV and
only two solutions with two vanishing VEVs. These are
\renewcommand{\labelenumi}{(\alph{enumi})}
\begin{enumerate}
\item $\!\Phi_{12} = \Phi_3=0$, with $\det{M_D^{(3)}}=-4 F_1\oF_5 \phi_{45}\ophi_{45}\oPhi_{12}$ and massless doublets
\begin{align}
H &= \phi_{45} H_1 - \oPhi_{31} H_{45}\label{H}\,,\\
\bar{H} &=\ophi_{45} \oH_2 - \oPhi_{23} \oH_{45}\,.\label{Hb}
\end{align}
\item $\!\Phi_{12} = \oPhi_{31}=0$, with
$\det{M_D^{(3)}}=-4 F_1\oF_5\left(
\Phi_{23}\Phi_{31}\Phi_3
+\phi_{45}\ophi_{45}\oPhi_{12}
\right)$ and massless doublets
\begin{align}
H &\!=\! H_1\,,\\
\oH &\!=\!\Phi_{23}\oPhi_{23}\Phi_3 \oH_1
\!-\! \left(\Phi_{31}\Phi_{23}\Phi_3\!+\!\oPhi_{12} \phi_{45}\ophi_{45}\right) \oH_2
\!-\! \oPhi_{12}\oPhi_{23} \Phi_{3} \oH_3
 \!+\! \oPhi_{12}\oPhi_{23} \phi_{45} \oH_{45}
\end{align}
\end{enumerate}

\section{Solution of the F/D-flatness conditions}
\label{FDsol}
In supersymmetric string models build in the framework of the free fermionic formulation, many features of low energy phenomenology are specified by the VEVs of Standard Model singlet fields. These in turn, are subject to 
%a considerable number of 
non-trivial F/D flatness constraints dictated by $N=1$ supersymmetry.  For the flipped string model  \cite{Antoniadis:1989zy} 
the solution and phenomenological consequences of F/D flatness conditions have been thoroughly studied in the past
\cite{Antoniadis:1989zy, Lopez:1989fb, Lopez:1990wt, Lopez:1991ac, Rizos:1990xn,  Antoniadis:1991fc}. In this section we re-examine solutions of  flatness equations taking into account: (i) compatibility with the recent results concerning string cosmology~\cite{Antoniadis:2020txn} as explained in Section \ref{scosmology} and (ii) the results of Section  \ref{doublet-triplet} regarding doublet-triplet splitting and in particular the requirement to generate triplet masses at tree-level in order to efficiently suppress proton decay.

For the flipped string model under consideration the full gauge group is $SU(5)\times{U(1)}\times{U(1)'}^4\times{SU(4)}\times{SO(10)}$
The D-flatness conditions associated with the ${U(1)'}^4$ gauge group factor
are of the form
\begin{align}
{\cal D}_I &= \sum_{i} q^i_I \left|\varphi_i\right|^2+\frac{1}{192\pi^2}{2\over\alpha'}\,{\rm Tr}Q_I=0\,, I=1,2,3,A\,,\label{DFI}
\end{align}
where %$g$ is the four dimensional gauge coupling, 
$\varphi_i$ denotes a field with charges $q^i_I$ under ${U(1)'}_I$
and  ${\rm Tr}{Q_1}={\rm Tr}{Q_2}={\rm Tr}{Q_3}=0$, ${\rm Tr}{Q_A}=180$ are the traces of the associated abelian group generators \cite{Antoniadis:1989zy, Cvetic:1998gv}. After some algebra Eqs. \eqref{DFI} can be unravelled by taking appropriate linear combinations \cite{Rizos:1990xn}
\begin{align}
%dflat1 in mathematica
&\left|\phi_{45}\right|^2  - \left|\overline{\phi}_{45}\right|^2 - \frac{1}{2}\left( \left|D_3\right|^2 + \left|T_3\right|^2\right)
+\frac{1}{2} \left|F_3\right|^2
 = 
\xi^2\,, \label{df1}\\
%dflat4 in mathematica
&\left|\phi_+\right|^2 - \left|\phi_-\right|^2 - \left|\overline{\phi}_+\right|^2 + \left|\overline{\phi}_-\right|^2 + \frac{1}{2}
\left(\left|D_3\right|^2  - \left|T_3\right|^2\right) - \frac{1}{2} \left|F_3\right|^2= \xi^2\,,\label{df2}\\
%dflat3 in mathematica
&\left|\Phi_{31}\right|^2 -\left|\overline{\Phi}_{31}\right|^2 
-\left|\Phi_{23}\right|^2 +\left|\overline{\Phi}_{23}\right|^2 
-\frac{1}{2}\left(\left|D_1\right|^2 + \left|D_2\right|^2 + \left|D_3\right|^2 + \left|D_4\right|^2 - \left|D_5\right|^2\right)\nonumber\\
&\ 
%dflat2 in mathematica
-\frac{1}{2} \left(\left|T_1\right|^2 + \left|T_2\right|^2 + \left|T_3\right|^2 - \left|T_4\right|^2 + \left|T_5\right|^2\right)
= 3\xi^2\,,\label{df3}
\\
&\left|\Phi_{23}\right|^2  - \left|\overline{\Phi}_{23}\right|^2
-\left|\Phi_{12}\right|^2  + \left|\overline{\Phi}_{12}\right|^2
+\frac{1}{2}\sum_{i=1}^4 \left(\left|\phi_i\right|^2-\left|\overline{\phi}_i\right|^2\right)
+ \left|\phi_+\right|^2 - \left|\overline{\phi}_+\right|^2 \nonumber\\
&\ +\frac{1}{2}\left(\left|D_1\right|^2 + \left|D_3\right|^2 + \left|D_4\right|^2\right)
+ \frac{1}{2}\left(\left|T_1\right|^2 - \left|T_4\right|^2 \right) -\frac{1}{2}
\left(\left|F_2\right|^2-\left|\oF_5\right|^2
\right)
= 0\,,
\label{df4}
\end{align}
where
\be\label{xi}
\xi^2 =\frac{1}{16\pi^2}{2\over\alpha'}\,.
\ee

The F-flatness equations are derived from the superpotential $W$ 
\begin{align}
{\cal F}_i = \frac{\partial W}{\partial \varphi_i}=0\,.
\end{align} 
At tree-level $W=W_3$ given in Eq. \eqref{wtree}. However, the superpotential 
receives additional contributions from  non-renormalisable (NR) terms  at higher orders.
These come from terms of the form $\varphi_1 \varphi_2 \varphi_3 \varphi^{N-3}, N>3$, where $\varphi^{N-3}$ denotes a product of $N-3$ field VEVs.
Apart from gauge invariance these terms are subject to intricate string selection rules 
ascribed to world-sheet superalgebra \cite{Lopez:1990wt,Rizos:1991bm}. Using a
computer program that successively applies all selection criteria we find 15 candidate NR superpotential couplings for $N=4$ and 256 couplings for $N=5$. 
However, for the sake of simplicity we start our analysis from the tree-level superpotential \eqref{wtree} and we will take into account higher order non-renormalisable contributions at a later stage. The tree-level F-flatness equations give
\begin{align}
\frac{\partial W_3}{\partial \Phi_{12}} &= \sum_{i=1}^4\phi_i^2+\Phi_{23}\Phi_{31} + \phi_+ \phi_-=0\,,
\label{fPhi12}\\
\frac{\partial W_3}{\partial \overline{\Phi}_{12}}  &= \sum_{i=1}^4\overline{\phi}_i^2+\overline{\Phi}_{23}\overline{\Phi}_{31}
+ \overline{\phi}_+ \overline{\phi}_-=0\,,\label{fPhib12}\\
\frac{\partial W_3}{\partial \Phi_{31}} &= \Phi_{12} \Phi_{23} +  D_2^2 + T_2^2 +T_5^2 =0\,,\\
\frac{\partial W_3}{\partial \oPhi_{31}} &= \overline{\Phi}_{12}\overline{\Phi}_{23} +  D_5^2 =0\,,\label{fPhib31}
\\
\frac{\partial W_3}{\partial \Phi_{23}} &= {\Phi}_{12}{\Phi}_{31}+ T_4^2 =0\,,\label{fPhi23}\\
\frac{\partial W_3}{\partial \oPhi_{23}} &= \overline{\Phi}_{12}\overline{\Phi}_{31} + D_1^2+D_4^2+T_1^2 =0\,,\\
\frac{\partial W_3}{\partial \Phi_{3}} &= \sum_{i=1}^4\phi_i\overline{\phi}_i + \phi_{45}\overline{\phi}_{45} + 
\phi_-\overline{\phi}_- + \phi_+\overline{\phi}_+ =0\,,\\
\frac{\partial W_3}{\partial \Phi_{4}}  &= \phi_2\overline{\phi}_1 + \phi_1\overline{\phi}_2 = 0\,,\label{fPhi4}\\
\frac{\partial W_3}{\partial \Phi_{5}}  &= \phi_4\overline{\phi}_3 + \phi_3\overline{\phi}_4 = 0\,,\label{fPhi5}
\\
\frac{\partial W_3}{\partial \phi_1}&= 2 \phi_1 \Phi_{12} + 
\Phi_3\overline{\phi}_1 + \Phi_4 \overline{\phi}_2 = 0\,,\label{fphi1}\\
\frac{\partial W_3}{\partial \phi_{2}} &=  2 \phi_2 \Phi_{12} + \Phi_4\overline{\phi}_1 + \Phi_3 \overline{\phi}_2  + T_4{\cdot}T_5 = 0\,,\\
\frac{\partial W_3}{\partial \phi_3} &= F_4 \oF_5 + 2 \phi_3 \Phi_{12} + \Phi_3\overline{\phi}_3 + \Phi_5 \overline{\phi}_4  = 0\,,\label{fphi3}\\
\frac{\partial W_3}{\partial \phi_4} &= 2\phi_4 \Phi_{12} + \Phi_5\overline{\phi}_3 + \Phi_3 \overline{\phi}_4 = 0\,,\label{fphi4}
\\
\frac{\partial W_3}{\partial \ophi_1}&= 2 \ophi_1 \oPhi_{12} + \Phi_3 {\phi}_1 + \Phi_4 {\phi}_2 = 0\,,\\
\frac{\partial W_3}{\partial \ophi_2} &=  2 \ophi_2 \oPhi_{12} + \Phi_4 {\phi}_1 + \Phi_3{\phi}_2   = 0\,,\label{fphib2}\\
\frac{\partial W_3}{\partial \ophi_3} &= 2 \ophi_3 \oPhi_{12} + \Phi_3 {\phi}_3 + \Phi_5 {\phi}_4  + D_4{\cdot}D_5= 0\,,
\label{d45}\\
\frac{\partial W_3}{\partial \ophi_4} &= 2\ophi_4 \oPhi_{12} + \Phi_5 {\phi}_3 + \Phi_3 {\phi}_4 = 0\,,\label{fphib4}\\
\frac{\partial W_3}{\partial \phi_+} &= \Phi_3 \ophi_+ + \Phi_{12} \phi_- = 0\,, \label{ep}\\
\frac{\partial W_3}{\partial \ophi_+} &= \Phi_3 \phi_+ + \oPhi_{12} \ophi_- = 0\,,\label{fphibp}\\
\frac{\partial W_3}{\partial \phi_-} &= \Phi_3 \ophi_- + \Phi_{12} \phi_+ = 0\,,\label{em}\\
\frac{\partial W_3}{\partial \ophi_-} &= \Phi_3 \phi_- + \oPhi_{12} \ophi_+ = 0\,,\label{fphibm}\\
\frac{\partial W_3}{\partial \phi_{45}} &= \Phi_3 \ophi_{45} = 0 \,,\label{fphi45}\\
\frac{\partial W_3}{\partial \ophi_{45}}&= \Phi_3 \phi_{45} = 0\,, \label{fphib45}\\
\frac{\partial W_3}{\partial F_4} &= \oF_5 \phi_3=0\label{fF4}\,,\\
\frac{\partial W_3}{\partial \oF_5} &= F_4 \phi_3=0\,,\label{fFb5}
\\
\frac{\partial W_3}{\partial T_1} &= 2 \overline{\Phi}_{23} T_1 =0\,,\label{fT1}\\  
\frac{\partial W_3}{\partial T_2} &=  2 \Phi_{31} T_2 =0\,,\label{fT2}\\
\frac{\partial W_3}{\partial T_4} &= 2 \Phi_{23} T_4 + \phi_2 T_5 =0\,,\\
\frac{\partial W_3}{\partial T_5} &=  \phi_2 T_4 + 2 \Phi_{31} T_5 = 0\label{fT5}\,,\\
\frac{\partial W_3}{\partial D_1} &= 2 D_1 \overline{\Phi}_{23} = 0\,,
\label{fD1}\\
\frac{\partial W_3}{\partial D_2}&= 2 D_2 {\Phi}_{31} = 0\,,\label{fD2}\\
\frac{\partial W_3}{\partial D_4} &= 2 D_4 \overline{\Phi}_{23} + D_5 \overline{\phi}_3/\sqrt{2} = 0\,,
\label{fD4}\\
\frac{\partial W_3}{\partial D_5} &= 2 D_5 \overline{\Phi}_{31} + D_4 \overline{\phi}_3/\sqrt{2} = 0\,.\hspace{3cm}
\label{fD5}
\end{align}

The massless spectrum of the string derived  flipped $SU(5)$  model comprises four fermion generations 
$F_\alpha, \of_\beta, \ell^c_\beta, \alpha=1,2,3,4, \beta=1,2,3,5$ and one anti-generation 
$\oF_5, f_4, \oell^c_4$.
The solution of the F/D-flatness equations depends on the choice of the flipped $SU(5)$ breaking Higgs fields $F, \oF$ and the assignment of the fermion generations. Following the discussion in Section \ref{doublet-triplet} we choose $F=A_1 F_1 + A_3 F_3 , \oF=\oF_5$ with $A_1^2+A_4^2=1$ and set
\begin{align}
F_2=F_4=0\,.
\label{sf1}
\end{align}
Combined with \eqref{df1}, this implies $\ophi_{45}\ne0$, which, in conjunction with \eqref{fphi45}, leads to $\Phi_3 =0\,$. 
This rejects the second solution of Section \ref{doublet-triplet} leaving us with
\begin{align}
\Phi_{12}=\Phi_3=0
\label{sf2}
\end{align}
and $\oPhi_{12}, \Phi_{31}, \oPhi_{31}, \Phi_{23}, \oPhi_{23}, \phi_{45}, \ophi_{45} \ne 0$ in order solve the doublet-triplet problem at tree-level.  Eq. \eqref{fPhi23} then implies
\begin{align}
T_4^2=0\,,
\label{sf3}
\end{align}
and \eqref{fF4} yields
\begin{align}
\phi_3=0\,
\label{sf4}
\end{align}
as $\oF_5\ne0\,$. To protect $\of_5$, associated in Section  
\ref{doublet-triplet}  with the 3rd fermion generation, from receiving 
tree-level mass via the superpotential term $f_4 \of_5 \ophi_2$ we have to 
impose also
\begin{align}
\ophi_2 = 0\,.
\label{sf5}
\end{align}
Next we solve Eqs. \eqref{fPhi4}, \eqref{fPhi5} by choosing
\begin{align}
\phi_2=\ophi_3 = 0\,
\label{sf6}
\end{align}
and Eqs. \eqref{fphi3}, \eqref{fphib2} using
\begin{align}
\Phi_4=\Phi_5=0\,.
\label{sf7}
\end{align}
Keeping in mind that $\Phi_{31}, \oPhi_{31} \ne 0$ Eqs. \eqref{fT2}, 
\eqref{fT5}, \eqref{fD2}, \eqref{fD5} lead to 
\begin{align}
T_2=T_5=D_2=D_5=0\,.
\label{T2T5D2D5}
\end{align}

The remaining equations can be solved after considering the aforementioned  
non-renormalisable corrections to the superpotential \eqref{wtree}.  Taking 
into account 
the full superpotential $W_5$ incorporating NR contributions up to order $N=5$ 
and
utilising the tree-level partial solution \eqref{sf1}-\eqref{T2T5D2D5}
 the F-flatness equations yield
\begin{align}
\frac{\partial W_5}{\partial \Phi_{12}} &= \phi_1^2 + \phi_4^2 + \phi_+ 
\phi_-+\Phi_{23}\Phi_{31}=0\,,
\label{fPhi12m}\\
\frac{\partial W_5}{\partial \overline{\Phi}_{12}}  &= \ophi_1^2+\ophi_4^2
+ \overline{\phi}_+ \overline{\phi}_-+\overline{\Phi}_{23}\overline{\Phi}_{31}
+\left\{F_1^2 \oF_5^2\right\}
=0\,,\label{fPhib125}\\
\frac{\partial W_5}{\partial \Phi_{31}} &=
\left\{\left(\ophi_1^2+\ophi_4^2
+ \overline{\phi}_+ 
\overline{\phi}_-\right)\left(D_1^2+D_4^2+T_1^2\right)\right\}
 =0\,,
 \label{fPhi315}
 \\
\frac{\partial W_5}{\partial \oPhi_{31}} &= \oPhi_{12}\oPhi_{23} 
=0\,,\label{fPhib315}\\
\frac{\partial W_5}{\partial \oPhi_{23}} &= \oPhi_{12} \oPhi_{31} + 
D_1^2+D_4^2+T_1^2 =0\,,
\label{fPhib235}\\
\frac{\partial W_5}{\partial \Phi_{3}} &=  \phi_1\ophi_1+\phi_4\ophi_4 + 
\phi_-\overline{\phi}_- + \phi_+\overline{\phi}_+ + 
\phi_{45}\overline{\phi}_{45} =0\,,\\
\frac{\partial W_5}{\partial F_1} &= \left\{F_1 \oF_5^2 \oPhi_{12}\right\} 
=0\,,\\
\frac{\partial W_5}{\partial \oF_5} &= \left\{\oF_5 F_1^2 \oPhi_{12}\right\} 
=0\,,\\
\frac{\partial W_5}{\partial \phi_3} &= \left\{\left(F_1 
\oF_5\right)\left(T_1{\cdot}T_4\right)\right\} =0\,,\\
\frac{\partial W_5}{\partial \ophi_1} &= 2\ophi_1 \left[\oPhi_{12} + 
\left\{\Phi_{31}\left(D_1^2+D_4^2+T_1^2\right)\right\} \right] = 0\,,
\label{ophi15}
\\
\frac{\partial W_5}{\partial \ophi_4} &= 2\ophi_4 \left[\oPhi_{12} + 
\left\{\Phi_{31}\left(D_1^2+D_4^2+T_1^2\right)\right\} \right] = 0\,,
\label{ophi45}\\
\frac{\partial W_5}{\partial \ophi_+} &= \ophi_- \left[\oPhi_{12} + 
\left\{\Phi_{31}\left(D_1^2+D_4^2+T_1^2\right)\right\} \right] = 0\,,
\label{ophip5}\\
\frac{\partial W_5}{\partial \ophi_-} &= \ophi_+ \left[\oPhi_{12} + 
\left\{\Phi_{31}\left(D_1^2+D_4^2+T_1^2\right)\right\} \right] = 0\,,
\label{ophim5}
\\
\frac{\partial W_5}{\partial T_1} &= 2 T_1  
\left[\oPhi_{23}+\left\{\Phi_{31}\left(\ophi_1^2+\ophi_4^2+\ophi_+ 
\ophi_-\right)\right\}\right] = 0\,,
\label{T15}
\\
\frac{\partial W_5}{\partial T_4}&=   2 T_4 \left[
\Phi_{23} + \left\{\oPhi_{31} \left(\phi_1^2+\phi_4^2+\phi_+ 
\phi_-\right)\right\}
\right] = 0\,,
\label{T45}
\\ 
\frac{\partial W_5}{\partial D_1} &= 2 D_1  
\left[\oPhi_{23}+\left\{\Phi_{31}\left(\ophi_1^2+\ophi_4^2+\ophi_+ 
\ophi_-\right)\right\}\right] = 0\,,
\label{D15}
\\
\frac{\partial W_5}{\partial D_4} &= 2 D_4  
\left[\oPhi_{23}+\left\{\Phi_{31}\left(\ophi_1^2+\ophi_4^2+\ophi_+ 
\ophi_-\right)\right\}\right] = 0\,,
\label{D45}
\\
\frac{\partial W_5}{\partial D_5} &= \left\{D_3 \left(F_3\oF_5\right)\right\} = 
0\,,
\label{D55}
\end{align}
where we have  omitted order one numerical coefficients of the NR terms  (terms 
in curly brackets). 

Higher order superpotential terms lead to additional constraints on the SM 
singlet VEVs in order to assure compatibility with low
energy phenomenology. First, superpotential terms of the form $F_1 \of_j \oh_k 
\varphi^n, j=1,2,3,5, k=1,2,3,45$ could induce mixings of leptons with Higgs 
doublets. The effective superpotential $W_5$ comprises three such terms 
\begin{align}
 F_1 \of_1 \oh_{45} \phi_1(1+\Phi_2) + F_1 \of_5 \oh_{45} 
 \left(T_1{\cdot}T_4\right)\,,
\end{align}
where we dropped curly brackets (denoting omission of numerical coefficients), 
for simplicity.
To eliminate them we set
\begin{align}
\phi_1=0\ ,\ T_1{\cdot}T_4 = 0\,.
\end{align}
Second, the fifth order superpotential $W_5$ includes two terms that provide 
mass for the surplus pairs of fermions $f,\of$ and  $\ell^c, \oell^c$
\begin{align}\label{fbarlc34}
\left(f_4 \of_3 + \oell^c_4 \ell^c_3 \right) \left(T_3\cdot{T_4}\right)\,,
\end{align}
as long as
\begin{align}
T_3\cdot{T_4}\ne0\,.
\end{align}
%Finally, for reasons explained earlier in this section we will choose 
%$F_3\ne0$, thus we set 
%\begin{align}
%D_3 =0
%\end{align}
%to satisfy \eqref{D55}.

Summarizing the VEV assignments above, we have
\begin{gather}
F_2=F_4=0\,,\nonumber\\
\Phi_{12}=\Phi_3=\Phi_4=\Phi_5=\phi_1=\phi_2=\phi_3=\ophi_2=\ophi_3=0\,,\\
T_2=T_5=D_2=D_5=T_4^2=T_1{\cdot}T_4=0\,,\nonumber
\end{gather}
together with
\begin{align}
\oPhi_{12}, \Phi_{31}, \oPhi_{31}, \Phi_{23}, \oPhi_{23}, \phi_{45}, 
T_3\cdot{T_4}\ne0\,.
\end{align}

Let us now proceed with the solution of the remaining equations. Notice that the system of D/F-flatness constraints contains a single fixed parameter $\xi$ \eqref{xi}, which is about 0.1 %of order of $10^{-1}-10^{-2}$ 
in string units. We can thus attempt to find a perturbative solution expressing all remaining VEVs  as power series in $\xi$. Assuming that (to leading order)
$\Phi_{31}, \oPhi_{31}, \phi_{45},\phi_4,\phi_+,\phi_-, \ophi_1, \ophi_\pm\sim \xi$, 
%$\phi_4^2+\phi_+ \phi_-%, \ophi_1^2+\ophi_+ \ophi_-\sim\xi^2$,  
$F_1 \oF_5 \sim \xi^3$, Eq. \eqref{T45} can be solved with 
respect to $\Phi_{23}$
\begin{align}
\Phi_{23}  = - \left\{\oPhi_{31} \left(\phi_4^2+\phi_+ \phi_-\right)\right\} \sim \xi^3\,,
\end{align}
Eqs. \eqref{T15}, \eqref{D15}, \eqref{D45} with respect to  $\oPhi_{23}$
\begin{align}
\oPhi_{23} = - \left\{\Phi_{31}\left(\ophi_1^2+\ophi_4^2+\ophi_+ 
\ophi_-\right)\right\} \sim \xi^3
\,,
\label{ffsol1}
\end{align}
and Eqs. \eqref{ophi15}-\eqref{ophim5} with respect to $\oPhi_{12}$
\begin{align}
\oPhi_{12} = - \left\{\Phi_{31}\left(D_1^2+D_4^2+T_1^2\right)\right\} \sim \xi^3\,.
\label{ffsol2}
\end{align}
The solutions \eqref{ffsol1}, \eqref{ffsol2}  need further clarification, since 
both involve three different equations that should be compatible 
(\eqref{ophip5} and \eqref{ophim5} come from the same superpotential term). 
Actually, their validity depends on the numerical factors of the fifth order 
terms  that we have omitted (terms in curly brackets). A detailed analysis of 
Eqs. \eqref{fPhib125}-\eqref{fPhib235}, \eqref{ophi15}-\eqref{T15} and 
\eqref{D15},\eqref{D45}
% \cref{fPhib125,fPhi315,fPhib315,fPhib235,ophi15,ophip5,ophim5,T15,D15,D45}
taking into account all numerical coefficients is presented in the next section and Appendix \ref{appb} where we compute the F-flatness solutions of a generic superpotential of the form
\begin{align}
w&=\oPhi_{23}\oPhi_{31}\oPhi_{12} +
\oPhi_{12} \left(\ophi_1^2+\ophi_2^2+\ophi_4^2+\ophi_+\ophi_-\right)
\nonumber\\
&+\left(D_1^2+D_4^2+T_1^2\right)\oPhi_{23}+\left(\alpha_1 \ophi_1^2+\alpha_2 
\ophi_2^2+\alpha_3 \ophi_4^2+\alpha_4\ophi_+ \ophi_-\right) D_1^2 \Phi_{31} 
\label{NRTeqs}%\nonumber
\\
&+\left(\!\beta_1 \ophi_1^2\!+\!\beta_2 \ophi_2^2\!+\!\beta_3\ophi_4^2\!+\!\beta_4 
\ophi_+\ophi_-\!\right)\! D_4^2 \Phi_{31} \!+\! \left(\gamma_1 
\ophi_1^2+\gamma_2\ophi_2^2+\gamma_3\ophi_4^2+\gamma_4 \ophi_+\ophi_-\right)\!
T_1^2 \Phi_{31}.\nonumber
\end{align}
As shown in Section \ref{NRTcomputation} an explicit calculation of the related 
fifth order superpotential couplings supports the following coupling relations
\begin{align}
-\frac{\alpha_1}{3}=-\frac{\gamma_1}{3}=-\frac{\beta_1}{2}=\frac{\beta_3}{2}=
\frac{\beta_4}{2}=\alpha_4\,,\\
\alpha_2=\alpha_3=\gamma_2=\gamma_3=\gamma_4=
\alpha_4\ ,\ \beta_2=0\,.
\end{align}
These in turn, following the analysis of Appendix \ref{appb}, lead to the subsequent
solution of the associated F-flatness equations to order $\xi^4$
%\begin{align}
%\ophi_1^2&=-\ophi_4^2-\ophi_+\ophi_-+4\alpha_4\Phi_{31}\oPhi_{31}\left[\ophi_4^2+\ophi_+
% \ophi_-\right]+{\cal O}\left(\xi\right)^5 \,,\\
%T_1^2&=-D_1^2-D_4^2+\alpha_4 \Phi_{31}\oPhi_{31} D_4^2+{\cal 
%O}\left(\xi\right)^5\,,\\
%\oPhi_{23}&=-4\alpha_4 \Phi_{31}\left[\ophi_4^2+\ophi_+ \ophi_-\right] +{\cal 
%O}\left(\xi\right)^5\,,\\
%\oPhi_{12} &= -\alpha_4 \Phi_{31} D_4^2+{\cal O}\left(\xi\right)^5\,.
%\end{align}
\begin{align}
\oPhi_{12} &= - \alpha_4 D_4^2 \Phi_{31}\\
T_1^2&=-D_1^2-D_4^2 - \oPhi_{31}\oPhi_{12} \,,
\\
\Phi_{23}  &= - \left\{\oPhi_{31} \left(\phi_4^2+\phi_+ \phi_-\right)\right\}\\
\oPhi_{23}&=-2\Phi_{31}\alpha_4\left(-\ophi_1^2 + \ophi_4^2+  \ophi_+ \ophi_- 
\right)\label{Phibar23}\\
\ophi_1^2 &=  -\ophi_4^2-\ophi_+ \ophi_--\oPhi_{23} \oPhi_{31}\,,
\end{align}
where in \eqref{Phibar23} we used the exact coefficients from the computation.

The remaining F and D-flatness equations  to order $\xi^4$ give
\begin{align}
\phi_4^2 + \phi_+ \phi_- &= 0\,,\\
\phi_4\ophi_4+\phi_-\overline{\phi}_- + \phi_+\overline{\phi}_+ + 
\phi_{45}\overline{\phi}_{45} &=0\,,\\
\left\{D_3 \left(F_3\oF_5\right)\right\} &= 0\label{ed3}
\,,
\end{align}
and 
\begin{align}
%dflat1 in mathematica
&\left|\phi_{45}\right|^2  - \left|\overline{\phi}_{45}\right|^2 +
 \frac{1}{2}\left( 
 \left|F_3\right|^2
 -\left|D_3\right|^2 - \left|T_3\right|^2\right)
 = 
\xi^2\,, \label{dff1}\\
%dflat4 in mathematica
&\left|\phi_+\right|^2 - \left|\phi_-\right|^2 - \left|\overline{\phi}_+\right|^2 + \left|\overline{\phi}_-\right|^2 + \frac{1}{2}
\left(\left|D_3\right|^2  - \left|T_3\right|^2-\left|F_3\right|^2\right) = \xi^2\,,\label{dff2}\\
%dflat3 in mathematica
&\left|\Phi_{31}\right|^2 \!-\! \left|\overline{\Phi}_{31}\right|^2 
\!-\!\frac{1}{2}\left(\left|D_1\right|^2  \!+\! \left|D_3\right|^2 \!+\! \left|D_4\right|^2\right)
%dflat2 in mathematica
\!-\!\frac{1}{2} \left(\left|T_1\right|^2  \!+\! \left|T_3\right|^2 \!-\! \left|T_4\right|^2 \right)
= 3\xi^2\,,\label{dff3}
\\
&
%dflat2 in Mathematica
\frac{1}{2} 
\left(\left|\phi_4\right|^2-\left|\ophi_1\right|^2-\left|\ophi_4\right|^2\right)
+ \left|\phi_+\right|^2 - \left|\overline{\phi}_+\right|^2 +\frac{1}{2}\left(\left|D_1\right|^2 +
\left|D_3\right|^2 + \left|D_4\right|^2\right) \nonumber\\
&
\ \ + \frac{1}{2}\left(\left|T_1\right|^2 - \left|T_4\right|^2 -\left|\oF_5\right|^2\right) 
= 0\,,
\label{dff4}
\\
&\left|A_1 F_1\right|^2 + \left|A_3 F_3\right|^2 = \left|\oF_5\right|^2\,.
\label{FFbar}
\end{align}
%respectively. 
In addition, we have the $SU(4)\simeq SO(6)$ and
 $SO(10)$ D-flatness which can be cast in the form
\begin{align}
\sum_{i=1,4} D_i^\ast \tau^a D_i =0\ ,\ a=1,\dots,15\ ,
\label{dd4}\\
\sum_{i=1,3,4} T_i^\ast \lambda^A T_i =0\ ,\ A=1,\dots,45\ ,
\label{dd10}
\end{align}
where $\tau^a, \lambda^A$ are the $SO(6), SO(10)$ generators respectively. These can be solved by utilising an antisymmetric representation of the $SO(2n)$ group generators
\begin{align}
\left(M_{ab}\right)_{ij} = -i \left(\delta_{ai}\delta_{bj} - \delta_{aj}\delta_{bi}\right)
\ ;\ b>a=1,\dots,2n\ ;\ i,j= 1,\dots,2n\,.
\end{align}
In this way, Eqs. \eqref{dd4} can be easily solved by choosing the VEVs $D_1, D_3,D_4$ real. Although this is not applicable for Eqs. \eqref{dd10} because of the constraints $T_4^2=T_1{\cdot}T_4=0$, we can find other explicit solutions. For example, using  the ansatz
\begin{align}
T_1 &= \left(r e^{-i\theta}, i r e^{-i\theta},b,0,\dots,0\right)\,,\\
T_4 &= \left(a, i a,0,0,\dots,0\right)\,,\\
T_3 &= \left( i c, c , c+i e,0,\dots,0\right)\,,
\end{align}
Eqs. \eqref{dd10} reduce to three independent constraints solved by
\begin{align}
 r^2 = c^2 - a^2 \ ,\ e^2 = b^2 r^2/c^2- c^2\ ,\ 
\tan\theta=\frac{c}{e}\,,
\end{align}
with $a, b, c$ free real parameters. Moreover, this solution guarantees 
$T_4^2=T_1{\cdot}T_4=0$ and gives $T_1^2=b^2$ and $T_3{\cdot}T_4 = 2 i a c$, 
depending on the free parameters.

A natural question to ask at this point is whether higher order terms in the superpotential could destabilise our flatness solution. A definite answer to this  would require the calculation of the full superpotential to a rather high order, e.g. $N=10$, and the solution of the associated flatness equations, which is a very difficult task from the technical point of view. Here, we have shown that our 
perturbative solution is valid to order $\xi^4$ when taking into account 
superpotential terms up to and including $N=5$. Moreover, we have checked the 
$N=6$ contributions  and it turns out that the aforementioned flatness 
solution  holds to order $\xi^5$ provided $\Phi_2=0$, $D_3 (F_3 
\oF_5)\lesssim \xi^6$,
$T_3 T_4 \lesssim 
\xi^3$ and an additional condition relating the VEVs of $D_1, T_1$. 
This 
supports the perturbative validity of our solution at higher orders. The above 
VEV assignments that solve the F/D-flatness equations are summarised in the 
right panel of Table~\ref{tableofvevs}. 

\section{Computation of higher order superpotential terms} 
\label{NRTcomputation}
In the context of the free fermionic formulation of the heterotic superstring the 
 effective $N=1$ superpotential is fully calculable. Actually,
the trilinear coupling constants are fixed by conformal symmetry alone \cite{Belavin:1984vu};  the  non-vanishing  ones are of the form $k g_s$, where $g_s$ is the string coupling and $k\in\left\{1/{\sqrt{2}}, 1, \sqrt{2}\right\}$. The computation of the coupling constants of higher order NR  terms, $N>3$,  is in general more intricate as it entails the calculation of $N$-point correlation functions \cite{Kalara:1990fb}. Though suppressed  by inverse powers of the string scale, these terms turn out to play an important role in low energy phenomenology. For example, they can account for fermion mass hierarchies and provide intermediate scale masses for exotic states.  In our analysis they are also important in ensuring the existence 
of a particular solution of the F-flatness constraints as explained in Section \ref{FDsol} and Appendix \ref{appb}. For this purpose, we compute in this section, the coupling constants $\alpha_i,\beta_i,\gamma_i, i=1,4$ of the fifth order NR terms appearing in Eq.~\eqref{NRTeqs} 
\begin{align}
\!\!\!\left(\alpha_1 \ophi_1^2\!+\alpha_2 \ophi_2^2+\!\alpha_3 
\ophi_4^2\!+\!\alpha_4\ophi_+ \ophi_-\right)\! D_1^2 \Phi_{31} 
\!+\!\left(\beta_1 \ophi_1^2\!+\beta_2 \ophi_2^2\!+\beta_3 
\ophi_3^2\!+\!\beta_4 \ophi_+\ophi_-\right)\! D_4^2 \Phi_{31} \!\nn
\\+\! 
\left(\gamma_1 \ophi_1^2\!+\gamma_2 \ophi_2^2\!+\gamma_3 \ophi_4^2\!+\!\gamma_4 
\ophi_+\ophi_-\right)\! T_1^2 \Phi_{31} \,.
\end{align}
%where we dropped $\ophi_4$ which is of order $\xi^2$.

Following \cite{Kalara:1990fb} the coupling constant of a NR superpotential term of the form
\begin{align}
\int d^2\theta\,{\mathds X}_1\dots  {\mathds X}_N \ ,\ N\ge3\,
\end{align}
is proportional to the correlation function
\begin{align}
\langle{\Psi_1 \Psi_2 \Phi_3 \dots \Phi_N }\rangle\,,
\label{cfbb}
\end{align}
where $\Psi_1, \Psi_2$ are the fermionic components of the superfields ${\mathds X}_1, {\mathds X}_2$ respectively and $\Phi_i$ stand for the bosonic components of ${\mathds X}_i, i=3,\dots,N$. Evaluation of the correlator requires 
knowing the vertex operators of the associated fields which (in the 0-ghost picture) are world-sheet operators of conformal dimensions $(1,1)$. The vertex operators are
expressed in terms of the world-sheet degrees of freedom. These include
22 real left-moving fermions $\big\{\psi^\mu$, $\chi^{I},y^I,\omega^I, I=1,\dots,6\}$ together with 12 real $\big\{\overline{y}^I,\overline{\omega}^I, I=1,\dots,6\}$ and 16 complex $\big\{\bar{\psi}^{1,\dots,5},\bar{\eta}^{1,2,3},\bar{\phi}^{1,\dots,8}\big\}$ right-moving fermions.  In the case at hand, the world-sheet supercurrent takes the form
\begin{align}
T_F = \psi^\mu \partial X_\mu + i \sum_{I=1}^6 \chi^I y^I \omega^I \,.
\end{align}
Furthermore, the particular choice of basis vectors allows for the bosonisation of  the fermionic  fields $\chi^1,\dots,\chi^6$ as follows
\begin{align}
e^{\pm i S_{k,k+1}} = \frac{1}{\sqrt{2}} \left(\chi^k\pm i \chi^{k+1}\right)\ ;\ 
%e^{-i S_{k,k+1}} = \frac{1}{\sqrt{2}} \left(\chi^k-i \chi^{k+1}\right)\,,\ 
k=1,3,5
\,.
\end{align}
In terms of the bosonised fields, the supercurrent takes the form
\begin{align}
T_F = T^0_F + T^{-}_F + T^{+}_F\,,
\end{align}
where
\begin{align}
T^0_F &= \psi^\mu \partial X_\mu\,,\\
T^{-}_F & = \sum_{k=1,3,5} e^{- i S_{k,k+1}} \tau_{k,k+1}\,,\\
T^{+}_F & = -\left(T^{-}_F\right)^\ast\,,
\end{align}
and
\begin{align}
\tau_{k,\ell} = \frac{i}{\sqrt{2}}\left(y^k\omega^k + i y^\ell \omega^\ell\right)\,.
\end{align}
From the remaining real fermions only two pairs can be complexified, namely $\omega^2,\omega^3$ and $\overline{\omega}^2,\overline{\omega}^3$. The  rest of them can be grouped as nine left-right moving fermion pairs $\{y^I,\overline{y}^I\}, I=1,\dots,6$ and $\{\omega^I,\overline{\omega}^I\}, I=1,4,6$, ascribed to 
%an equal number of 
critical Ising models.

In this framework, a general vertex operator of the bosonic component of a chiral superfield in the canonical picture (ghost charge -1) and vanishing momentum is of the form
\begin{align}
V^B_{-1}(z)  &= e^{-c}(z) e^{i \alpha_1 S_{12}}(z) e^{i  \alpha_2 S_{34}}(z) e^{i  \alpha_3 S_{56}}(z) G(z,\oz) %e^{\frac{i}{2}K_\mu X^\mu}(z) e^{\frac{i}{2}K_\mu \overline{X}^\mu}(\overline{z})
\,,
\label{vb}
\end{align}
where %$X^\mu, \overline{X}^\mu$ stand for the space-time bosonic  world-sheet fields and 
$c$ stands for the ghost  field. Here, $\alpha_i \in \left\{0,\pm\frac{1}{2}, 
\pm1\right\}$ and  $G(z,\bar{z})$ stands for a conformal field with  dimensions 
$\left(\frac{1}{2}-\sum_{i=1}^3\frac{\alpha_i^2}{2},1\right)$ comprised of 
exponentials of the remaining bosonised fields and of primary Ising fields 
$\sigma_+(z,\overline{z}), \sigma_-(z,\overline{z}), f(z), 
\bar{f}(\overline{z})$,  corresponding to the order, disorder and the 
left/right fermion operators  respectively. The fermionic partner of \eqref{vb} 
in the canonical -1/2 picture takes the form
\begin{align}
V^F_{-1/2}(z)  &= e^{-c/2}(z) S_a e^{i (\alpha_1-1/2) S_{12}}(z) e^{i  (\alpha_2-1/2) S_{34}}(z) e^{i  (\alpha_3-1/2) S_{56}}(z)  %\nonumber\\ &\qquad \times  
G(z,\oz)  %e^{\frac{i}{2}K_\mu X^\mu}(z) e^{\frac{i}{2}K_\mu \overline{X}^\mu}(\overline{z})
\,,
\label{vf}
\end{align}
where $S_a$ represents a space-time spinor field.
In the computation of the correlator \eqref{cfbb} the bosonic fields $\Phi_4,\dots,\Phi_N$ have to be picture changed to the 0-ghost picture. This procedure can be carried out using the standard picture-changing formula
\begin{align}
V_0^B(z,\oz) = \lim_{w\to z} e^c(w) T_F(w) V_{-1}^B(z,\oz)
\label{pc}
\end{align}
and the relevant operator product expansions.

Let us now proceed with the computation of the correlator 
\begin{align}
\langle{{\psi^F_{1(-1/2)}}{\psi^F_{2(-1/2)}}\Phi_{31 (-1)}^B\varphi^B_{1(0)}\varphi^B_{2(0)}}\rangle\,,
\label{core}
\end{align}
where $\psi^F_1 \psi^F_2 \in \left\{\ophi_1^2, \ophi_2^2, \ophi_4^2, \ophi_+ 
\ophi_-\right\}$ and 
$\varphi^B_{1} \varphi^B_{2}\in \left\{D_1^2, D_4^2, T_1^2\right\}$. The vertex operators of the fields involved  are (in the canonical fermionic/bosonic picture)
\begin{align}
\ophi^F_{1(-1/2)}&= e^{-c/2} S_a e^{-\frac{i}{2} S_{56}} \sigma_+^{y_2} \sigma_+^{y_3} \oy^5 \sigma_+^{\omega_1} \sigma_-^{\omega_4} 
e^{-\frac{i}{2} \overline{H}_1} e^{+\frac{i}{2} \overline{H}_2}
%e^{\frac{i}{2}K_\mu X^\mu} e^{\frac{i}{2}K_\mu \overline{X}^\mu}
\,,\\
\ophi^F_{2(-1/2)}&= e^{-c/2} S_a e^{-\frac{i}{2} S_{56}} \sigma_+^{y_2} 
\sigma_+^{y_3} {\overline\omega}^5 \sigma_+^{\omega_1} \sigma_-^{\omega_4} 
e^{-\frac{i}{2} \overline{H}_1} e^{+\frac{i}{2} \overline{H}_2}
%e^{\frac{i}{2}K_\mu X^\mu} e^{\frac{i}{2}K_\mu \overline{X}^\mu}
\,,\\
\ophi^F_{4(-1/2)}&= e^{-c/2} S_a e^{-\frac{i}{2} S_{56}} \sigma_-^{y_2} 
\sigma_-^{y_3} {\overline\omega}^6 \sigma_-^{\omega_1} \sigma_+^{\omega_4} 
e^{-\frac{i}{2} \overline{H}_1} e^{+\frac{i}{2} \overline{H}_2}
%e^{\frac{i}{2}K_\mu X^\mu} e^{\frac{i}{2}K_\mu \overline{X}^\mu}
\,,\\
\ophi^F_{+(-1/2)}&= e^{-c/2} S_a e^{-\frac{i}{2} S_{56}} \sigma_+^{y_2} \sigma_-^{y_3} \sigma_+^{\omega_1} \sigma_+^{\omega_4} 
e^{-\frac{i}{2} \overline{H}_1} e^{+\frac{i}{2} \overline{H}_2} e^{-i \overline{H}_4}
%e^{\frac{i}{2}K_\mu X^\mu} e^{\frac{i}{2}K_\mu \overline{X}^\mu}
\,,\\
\ophi^F_{-(-1/2)}&= e^{-c/2} S_a e^{-\frac{i}{2} S_{56}} \sigma_+^{y_2} \sigma_-^{y_3} \sigma_+^{\omega_1} \sigma_+^{\omega_4} 
e^{-\frac{i}{2} \overline{H}_1} e^{\frac{i}{2}\overline{H}_2} e^{+i \overline{H}_4}
%e^{\frac{i}{2}K_\mu X^\mu} e^{\frac{i}{2}K_\mu \overline{X}^\mu}
\,,
\\
\Phi_{31 (-1)}^B &= e^{-c}e^{+i S_{34}} e^{+i \overline{H}_1} e^{-i \overline{H}_3}
%e^{\frac{i}{2}K_\mu X^\mu} e^{\frac{i}{2}K_\mu \overline{X}^\mu}
\\
D_{1 (-1)}^B &= e^{-c} e^{+\frac{i}{2} S_{34}}  e^{+\frac{i}{2} S_{56}}
\sigma_+^{y_3} \sigma_+^{y_4}\sigma_+^{y_5} \sigma_+^{y_6}
e^{-\frac{i}{2}\overline{H}_2} e^{+\frac{i}{2}\overline{H}_3}
e^{i J_{\mathbf{6}}{\cdot}W_{\mathbf{6}}}
%e^{\frac{i}{2}K_\mu X^\mu} e^{\frac{i}{2}K_\mu \overline{X}^\mu}
\,,
\\
D_{4 (-1)}^B &= e^{-c} e^{+\frac{i}{2} S_{34}}  e^{+\frac{i}{2} S_{56}}
\sigma_+^{y_3} \sigma_+^{y_6}\sigma_+^{\omega4} \sigma_+^{\omega_5}
e^{-\frac{i}{2}\overline{H}_2} e^{+\frac{i}{2}\overline{H}_3}
e^{i J_{\mathbf{6}}{\cdot}W_{\mathbf{6}}}
%e^{\frac{i}{2}K_\mu X^\mu} e^{\frac{i}{2}K_\mu \overline{X}^\mu}
\,,
\\
T_{1 (-1)}^B &= e^{-c} e^{+\frac{i}{2} S_{34}}  e^{+\frac{i}{2} S_{56}}
\sigma_+^{y_3} \sigma_-^{y_4}\sigma_-^{y_5} \sigma_+^{y_6}
e^{-\frac{i}{2}\overline{H}_2} e^{+\frac{i}{2}\overline{H}_3}
e^{i J_{\mathbf{10}}{\cdot}W_{\mathbf{10}}}
%e^{\frac{i}{2}K_\mu X^\mu} e^{\frac{i}{2}K_\mu \overline{X}^\mu} 
\,,
\end{align}
where $\sigma_\pm^f$ denotes the order/disorder operators of the Ising pair 
$\left\{f,\overline{f}\right\}$, $e^{i \overline{H}_i}, i=1,2,3$ and  $e^{i \overline{H}_4}$ stand for 
the bosonised fermions $\overline{\eta}^i, i=1,2,3 $ and
$\overline{\omega}^2, \overline{\omega}^3$ respectively; $e^{i J_{\mathbf{6}}{\cdot}W_{\mathbf{6}}}$ and $e^{i J_{\mathbf{10}}{\cdot}W_{\mathbf{10}}}$ represent the bosonised fermions 
ascribed to the vectorial representations of the hidden gauge group   $\mathbf{6}$  of $SO(6)\simeq SU(4)$ and $\mathbf{10}$ 
of $SO(10)$, respectively, where $J_{\mathbf{6}}, J_{\mathbf{10}}$ the associated bosonic fields and $W_{\mathbf{6}}, W_{\mathbf{10}}$ the corresponding charges. Using \eqref{pc}, we can then derive the expressions for the 
picture-changed bosonic fields that contribute to the correlator \eqref{core}
\begin{align}
\!\!\!\!D_{1 (0)}^B &\!=\! \frac{i}{2}e^{-\frac{i}{2} S_{34}}  e^{+\frac{i}{2} S_{56}}
\bigl[
\sigma_-^{y_3} \sigma_+^{y_4}\sigma_+^{y_5} \sigma_+^{y_6} \omega^3
\!+\! i \sigma_+^{y_3} \sigma_-^{y_4}\sigma_+^{y_5} \sigma_+^{y_6} \omega^4
\bigl] %\nonumber\\ &\qquad \times
e^{-\frac{i}{2}\overline{H}_2} e^{+\frac{i}{2}\overline{H}_3}
e^{i J_{\mathbf{6}}{\cdot}W_{\mathbf{6}}}
%e^{\frac{i}{2}K_\mu X^\mu} e^{\frac{i}{2}K_\mu \overline{X}^\mu}\,,
\\
\!\!\!\!D_{4 (0)}^B &\!=\! \frac{i}{2}e^{-\frac{i}{2} S_{34}}  e^{+\frac{i}{2} S_{56}}
\bigl[
\sigma_-^{y_3} \sigma_+^{y_6}\sigma_+^{\omega_4} \sigma_+^{\omega_5} \omega^3
\!+\! i \sigma_+^{y_3} \sigma_+^{y_6}\sigma_-^{\omega_4} \sigma_+^{\omega_5} y^4
\bigl] %\nonumber\\ &\qquad \times
e^{-\frac{i}{2}\overline{H}_2} e^{+\frac{i}{2}\overline{H}_3}
e^{i J_{\mathbf{6}}{\cdot}W_{\mathbf{6}}}
%e^{\frac{i}{2}K_\mu X^\mu} e^{\frac{i}{2}K_\mu \overline{X}^\mu}\,,
\\
\!\!\!\!T_{1 (0)}^B &\!=\! \frac{i}{2}e^{-\frac{i}{2} S_{34}}  e^{+\frac{i}{2} S_{56}}
\bigl[
\sigma_-^{y_3} \sigma_-^{y_4}\sigma_-^{y_5} \sigma_+^{y_6} \omega^3
\!+\! i \sigma_+^{y_3} \sigma_+^{y_4}\sigma_-^{y_5} \sigma_+^{y_6} \omega^4
\bigl] %\nonumber\\ &\qquad \times
e^{-\frac{i}{2}\overline{H}_2} e^{+\frac{i}{2}\overline{H}_3}
e^{i J_{\mathbf{10}}{\cdot}W_{\mathbf{10}}}
%e^{\frac{i}{2}K_\mu X^\mu} e^{\frac{i}{2}K_\mu \overline{X}^\mu}
\,.
\end{align}

Putting everything together, we get
\begin{align}
\alpha_1&=\langle{{\ophi^F_{1(-1/2)}}{\ophi^F_{1(-1/2)}}\Phi_{31 (-1)}^B D_{1(0)} D_{1(0)}}\rangle\nonumber\\
&=\frac{g_s^3\sqrt2}{\left(2\pi\right)^2} \prod_{i=1}^5\int d^2z_i\,\,
\langle{e^{-c/2}(1)}{e^{-c/2}(2)}{e^{-c}(3)}\rangle
\langle{S_a(1) S_b(2)}\rangle\nonumber\\
&\quad\times
\langle{e^{i S_{34}}(3)e^{-\frac{i}{2} S_{34}}(4)e^{-\frac{i}{2} S_{34}}(z_5)}\rangle
\langle{e^{-\frac{i}{2} S_{56}}(1) e^{-\frac{i}{2} S_{56}}(2)e^{+\frac{i}{2} S_{56}}(4)e^{+\frac{i}{2} S_{56}}(5)}\rangle\nonumber\\
&\quad\times
\langle{e^{-\frac{i}{2}\overline{H}_1}(1)e^{-\frac{i}{2}\overline{H}_1}(2) e^{+i \overline{H}_1}(3)}\rangle
\langle{e^{+\frac{i}{2}\overline{H}_2}(1)e^{+\frac{i}{2}\overline{H}_2}(2)
e^{-\frac{i}{2}\overline{H}_2}(4) e^{-\frac{i}{2}\overline{H}_2}(5) }\rangle\nonumber\\
&\quad\times
\langle{
e^{-\i\overline{H}_3}(1) e^{+\frac{i}{2}\overline{H}_3}(4) e^{+\frac{i}{2}\overline{H}_3}(5)  
}\rangle \times \Sigma(1,2,5,6)\times
\langle{e^{i J_{\mathbf{6}}{\cdot}W_{\mathbf{6}}}(5)e^{i J_{\mathbf{10}}{\cdot}W_{\mathbf{10}}}(6)}\rangle\nonumber\\
&\qquad\quad
\langle{e^{i J_{\mathbf{10}}{\cdot}W_{\mathbf{10}}}(5)e^{i J_{\mathbf{10}}{\cdot}W_{\mathbf{10}}}(6)}\rangle
%\times\prod_{i=1}^6\langle{
%e^{\frac{i}{2}K_i{\cdot}X_i}e^{\frac{i}{2}K_i{\cdot}\overline{X}_i}
%}\rangle
\label{a1a1}
\,,
\end{align}
where
\begin{align}
 \Sigma(1,2,5,6) =-\frac{1}{4}\left(\Sigma_1-\Sigma_2\right)\,,
\end{align}
with
\begin{align}
\Sigma_1 &=
\langle{\sigma_+^{y_2}(1)\sigma_+^{y_2}(2)}\rangle
\langle{\sigma_+^{\omega_1}(1)\sigma_+^{\omega_1}(2)}\rangle
\langle{\sigma_+^{y_4}(5)\sigma_+^{y_4}(6)}\rangle
\langle{\sigma_+^{y_6}(5)\sigma_+^{y_6}(6)}\rangle\,,
\nonumber\\
&\quad\times
\langle{\sigma_+^{y_3}(1)\sigma_+^{y_3}(2)\sigma_-^{y_3}(5)\sigma_-^{y_3}(6)}\rangle
\langle{\oy^5(1)\oy^5(2)\sigma_+^{y_5}(5)\sigma_+^{y_5}(6)}\rangle
\langle{\sigma_-^{\omega_4}(1)\sigma_-^{\omega_4}(2)}\rangle
\langle{\omega^3(5)\omega^3(6)}\rangle
\\
\Sigma_2 &=
\langle{\sigma_+^{y_2}(1)\sigma_+^{y_2}(2)}\rangle
\langle{\sigma_+^{\omega_1}(1)\sigma_+^{\omega_1}(2)}\rangle
\langle{\sigma_-^{y_4}(5)\sigma_-^{y_4}(6)}\rangle
\langle{\sigma_+^{y_6}(5)\sigma_+^{y_6}(6)}\rangle
\nonumber\\
&\quad\times
\langle{\sigma_+^{y_3}(1)\sigma_+^{y_3}(2)\sigma_+^{y_3}(5)\sigma_+^{y_3}(6)}\rangle
\langle{\oy^5(1)\oy^5(2)\sigma_+^{y_5}(5)\sigma_+^{y_5}(6)}\rangle
\langle{\sigma_-^{\omega_4}(1)\sigma_-^{\omega_4}(2)\omega^4(5)\omega^4(6)}\rangle\,.
\end{align}
Substituting the conformal correlators from Appendix \ref{appcoup}, %and assuming vanishing external momenta 
we get
\begin{align}
\alpha_1 = - \frac{g_s^3\sqrt2}{4\left(2\pi\right)^2}\, I_{\alpha_1}\,,
\end{align}
with
\begin{align}
 I_{\alpha_1} =
\int d^2 w \int d^2 z \left[K_1^{\alpha_1}(w,z) - K_2^{\alpha_1}(w,z)\right]\,.
\label{a1kk}
\end{align}
Here
\begin{align}
K_1^{\alpha_1}(w,z)&=
\frac{1}{2\sqrt2}
\left|w\right|^{-1}\left|1-w\right|^{-1}\left|w-z\right|^{-1}|z|^{-2}|1-z|^{-3/4}\nonumber\\
&\quad\times
\left(1-|z|+|1-z|\right)^{1/2} \frac{2-\bar{z}}{\sqrt{1-\bar{z}}}\,,\\
K_2^{\alpha_1}(w,z)&=
\frac{1}{4\sqrt2}
\left|w\right|^{-1}\left|1-w\right|^{-1}\left|w-z\right|^{-1} |z|^{-2}|1-z|^{-7/4}\nonumber\\
&\quad\times
 \left(1+|z|+|1-z|\right)^{1/2} |2-z|^2\,,
\end{align}
where we have utilised conformal invariance to set $z_1=\infty\,, z_2=1\,, z_3=w$, $z_4=z\,, z_5=0$.
It should be emphasised that both $K_1^{\alpha_1}$ and $K_2^{\alpha_1}$ exhibit a pole at $z=0$, however, the
poles cancel exactly in the expression of Eq. \eqref{a1kk} ensuring the convergence of the integral. Next, we change variables $z\to1-z, w\to1-w$ in $K_1^{\alpha_1}(w,z)$ 
\begin{align}
K_1^{\alpha_1}(w,z)&=
\frac{1}{2\sqrt2}
\left|w\right|^{-1}\left|1-w\right|^{-1}\left|w-z\right|^{-1}|z|^{-3/4}|1-z|^{-2}
\left(1-|z|+|1-z|\right)^{1/2}\nonumber\\
&\times\left(\sqrt{z}+\sqrt{\bar{z}}|z|\right)
\end{align}
and then replace $K_1^{\alpha_1}(w,z)$ with 
$(K_1^{\alpha_1}(w,z)+K_1^{\alpha_1}(\bar{w},\bar{z}))/2$ in \eqref{a1kk} that enables us to recast $K_1^{\alpha_1}$ in a purely real form:
\begin{align}
K_1^{\alpha_1}(w,z)&=
\frac{1}{2\sqrt2}
\left|w\right|^{-1}\left|1-w\right|^{-1}\left|w-z\right|^{-1}|z|^{-3/4}|1-z|^{-2}
\left(1-|z|+|1-z|\right)^{1/2}\nonumber\\
&\times\left(1+|z|\right) \text{Re}\sqrt{z}\,.
\end{align}

In a similar way, we find
\begin{align}
\alpha_4&=\langle{{\ophi^F_{+(-1/2)}}{\ophi^F_{-(-1/2)}}\Phi_{31 (-1)}^B D_{1(0)} D_{1(0)}}\rangle
=\frac{g_s^3\sqrt2}{\left(2\pi\right)^2}  I_{\alpha_4}
\label{a4a4}
\,,
\end{align}
with
\begin{align}
 I_{\alpha_4} =
\int d^2 w \int d^2 z \left[K_1^{\alpha_4}(w,z) - K_2^{\alpha_4}(w,z)\right]\,,
\label{a4kk}
\end{align}
and
\begin{align}
K_1^{\alpha_4}(w,z)&=
\frac{1}{\sqrt{2}}\left|w\right|^{-1} \left|1-w\right|^{-1}\left|w-z\right|^{-1} 
|z|^{-2}
\left|1-z\right|^{-3/4}\left(1+|z|+|1-z|\right)^{1/2}\,,\\
K_2^{\alpha_4}(w,z)&=\frac{1}{2\sqrt{2}}\left|w\right|^{-1}\left|1-w\right|^{-1}
\left|w-z\right|^{-1}\left|z\right|^{-2}\left|1-z\right|^{-3/4}
\left(1-|z|+|1-z|\right)^{1/2}\nonumber\\
&\times\left(1+|z|\right) \text{Re}\sqrt{z}\,.
\end{align}
Repeating the calculation for 
\begin{align}
\alpha_2&=\langle{{\ophi^F_{2(-1/2)}}{\ophi^F_{2(-1/2)}}\Phi_{31 (-1)}^B 
D_{1(0)} D_{1(0)}}\rangle\,,
\\
\alpha_3&=\langle{{\ophi^F_{4(-1/2)}}{\ophi^F_{4(-1/2)}}\Phi_{31 (-1)}^B D_{1(0)} D_{1(0)}}\rangle\,,
\end{align}
 we find
\begin{align}
\alpha_2=\alpha_3=\alpha_4\,. \label{a34}
\end{align}

The correlator functions  $\beta_1$,  $\beta_2$ and $\beta_4$ can be computed 
similarly
\begin{align}
\beta_1&=\langle{{\ophi^F_{1(-1/2)}}{\ophi^F_{1(-1/2)}}\Phi_{31 (-1)}^B D_{4(0)} D_{4(0)}}\rangle
=\frac{g_s^3\sqrt2}{\left(2\pi\right)^2}  I_{\beta_1}\,,\\
\beta_4&=\langle{{\ophi^F_{+(-1/2)}}{\ophi^F_{-(-1/2)}}\Phi_{31 (-1)}^B D_{4(0)} D_{4(0)}}\rangle
=\frac{g_s^3\sqrt2}{\left(2\pi\right)^2}  I_{\beta_4}
\label{b4b4}
\,,
\end{align}
where
\begin{align}
 I_{\beta_1} = -  I_{\beta_4} =
\int d^2 w \int d^2 z \left[K_1^{\beta_1}(w,z) - K_2^{\beta_1}(w,z)\right]\,,
\label{b1kk}
\end{align}
and
\begin{align}
K_1^{\beta_1}(w,z)&=
\frac{1}{2}\left|w\right|^{-1} \left|1-w\right|^{-1} \left|w-z\right|^{-1} \left|z\right|^{-2} \left|1-z\right|^{-1}
\left(|1-z|-|z|+1\right)
\,,\\
K_2^{\beta_1}(w,z)&=\frac{1}{2}\left|w\right|^{-1} \left|w-z\right|^{-1} \left|1-w\right|^{-1} \left|z\right|^{-2} \left|1-z\right|^{-1}
\left(|1-z|+|z|+1\right)\,.
\end{align}
The two terms in the integrand are here nicely combined, cancelling explicitly the pole at $z=0$ and yielding 
\begin{align}
 I_{\beta_1} = -  I_{\beta_4} =
-\int d^2 w \int d^2 z
\left|w\right|^{-1} \left|1-w\right|^{-1} \left|w-z\right|^{-1}\left|z\right|^{-1}\left|1-z\right|^{-1}\,.
\label{b1kkp}
\end{align}
Repeating for 
\begin{align}
\beta_2 = \langle{{\ophi^F_{2(-1/2)}}{\ophi^F_{2(-1/2)}}\Phi_{31 (-1)}^B 
D_{4(0)} D_{4(0)}}\rangle\,,\\
\beta_3 = \langle{{\ophi^F_{4(-1/2)}}{\ophi^F_{4(-1/2)}}\Phi_{31 (-1)}^B 
D_{4(0)} D_{4(0)}}\rangle\,,
\end{align}
we obtain
\begin{align}
\beta_2=0 \ ,\ 
\beta_3=\beta_4. \label{b3b4}
\end{align}
Finally, an analysis of the $\gamma_1,  \gamma_4$ correlators gives
\begin{align}
\gamma_1&=\langle{{\ophi^F_{1(-1/2)}}{\ophi^F_{1(-1/2)}}\Phi_{31 (-1)}^B T_{1(0)} D_{4(0)}}\rangle
=\frac{g_s^3\sqrt2}{\left(2\pi\right)^2}  I_{\beta_1}  =\frac{g_s^3\sqrt2}{\left(2\pi\right)^2}  I_{\alpha_1}= \alpha_1 
\label{g1g1}\,,\\
\gamma_4&=\langle{{\ophi^F_{1(-1/2)}}{\ophi^F_{1(-1/2)}}\Phi_{31 (-1)}^B T_{1(0)} D_{4(0)}}\rangle
=\frac{g_s^3\sqrt2}{\left(2\pi\right)^2}  I_{\beta_4}  =\frac{g_s^3\sqrt2}{\left(2\pi\right)^2}  I_{\alpha_4} = \alpha_4
\label{g4g4}
\,.
\end{align}
In the same way we get
\begin{align}
\gamma_2&=\langle{{\ophi^F_{2(-1/2)}}{\ophi^F_{2(-1/2)}}\Phi_{31 (-1)}^B 
T_{1(0)} D_{4(0)}}\rangle = \gamma_4 = \alpha_4\,,
\\
\gamma_3&=\langle{{\ophi^F_{4(-1/2)}}{\ophi^F_{4(-1/2)}}\Phi_{31 (-1)}^B 
T_{1(0)} D_{4(0)}}\rangle = \gamma_4 = \alpha_4\,. \label{g3g4}
\end{align}

Summarising, the couplings under consideration involve three independent integrals $I_{\alpha_1}, I_{\alpha_4}$ and $I_{\beta_1}$ defined in \eqref{a1kk}, \eqref{a4kk} and  \eqref{b1kk}, respectively. We can calculate them numerically after changing to  polar coordinates $w=\left|w\right| e^{iu}, z=\left|z\right| e^{iv}$. Using Mathematica we get
\begin{align}
I_{\alpha_1} = -567.20\ ,\ 
% myintegrate[idka1radial,{5}]=-567.19927746908518979
I_{\alpha_4} = 189.07\ ,\ 
% myintegrate[idka4radial,{3}]=189.04716820589847878
% myintegrate[idka4radial,{4}]=189.06997152683023041
% myintegrate[idka4radial,{5}]=189.07245497955472721,
I_{\beta_1} = -378.14
% myintegrate[dkb1radial, {4}]= -378.13646058346487757
% myintegrate[dkb1radial, {5}]= -378.14492641514871316
\end{align}
with an error of the order of 0.1\%. 
This supports the relations
\begin{align}
I_{\alpha_1} = - 3 I_{\alpha_4}\ ,\ I_{\beta_1} = - 2 I_{\alpha_4}\,,
\end{align}
yielding 
\begin{align}
\alpha_1 = -3\alpha_4\ ,\ \beta_1 = 3 \beta_4\,.
\end{align}
When combined with Eqs. \eqref{a34}, \eqref{b3b4},\eqref{g1g1}, \eqref{g4g4}, \eqref{g3g4} the last equation gives a set of coupling relations 
\begin{align}
-\frac{\alpha_1}{3} = -\frac{\gamma_1}{3} = -\frac{\beta_1}{2}= 
\frac{\beta_3}{2} = \frac{\beta_4}{2} = \gamma_3=\gamma_4 = \alpha_3=\alpha_4\,,
\\
\alpha_2 = \gamma_2 = \alpha_4\ ,\ \beta_2=0\,.
\end{align}

\section{Fermion masses and proton decay}
In this section we study the phenomenological consequences of the flatness solution of Section~\ref{FDsol}. Of key importance to low energy phenomenology are the surviving SM Higgs doublets. At tree-level these are given in Eqs. \eqref{H}, \eqref{Hb}. However, both the doublet and the triplet mass matrices receive additional contributions from NR terms of the form
$h_i \oh_{j} \varphi^{N-3}\,,i,j=1,2,3,45$, where $\varphi^{N-3}$ is a combination of  VEVs, arising from a NR term of order $N>3$. A comprehensive computer search up to and including $N=7$ order yields numerous NR terms of this form. However, after applying the flatness solution of Section \ref{FDsol} with the additional simplifying assumptions
\begin{align}
\Phi_1 = \Phi_2=0\, ,\quad \ophi_{45}\simlt\xi^2\,,
\end{align}
the doublet mass matrix reduces to
\begin{align}
M_H^{(7)} = \bordermatrix{
		~&H_1&H_2&H_3&H_{45}\cr
		{\oH}_1&0&\oPhi_{12}&\Phi_{31}&0\cr
		{\oH}_2&0&{A^{(7)}}&\oPhi_{23}&0\cr
		{\oH}_3&\oPhi_{31}&\Phi_{23}&0&\phi_{45}\cr
		{\oH}_{45}&0&\overline{B}^{(5)}&\ophi_{45}&0\cr
	} + {\cal O}\left(\xi^6\right)\,,
\end{align}
where 
\begin{align}
\overline{B}^{(5)} &= \left\{\ophi_{45}\left(D_1^2+D_4^2+T_1^2\right)\right\}\sim \xi^2\ophi_{45}\,,\\
A^{(7)} &= \left\{\Phi_{31} \left(\ophi_1^2+\ophi_4^2+\ophi_+\ophi_-\right)
\left(D_1^2+D_4^2+T_1^2\right)\right\} \sim \xi^5\,.
\end{align}
Diagonalising we get a massless doublet pair
\begin{align}\label{Hd}
H&\propto\phi_{45} H_1 -\oPhi_{31} H_{45}\,,\\
\oH&\propto \left(\overline{B}^{(5)}\oPhi_{23}\!-\!A^{(7)} \ophi_{45}\right)\! \oH_1 \!+\! 
{\left(\oPhi_{12} \ophi_{45}\!-\!\overline{B}^{(5)} \Phi_{31} \right)\!\oH_2} \!+\!
\left(A^{(7)}\Phi_{31}\!-\! \oPhi_{12}\oPhi_{23} \right)\!\oH_{45}\,
\label{Hu0}
\end{align}
In the last equation  the coefficient of $\oH_1$ is of order ${\cal O}(\xi^5\ophi_{45})$ while those of the components $\oH_2$ and $\oH_{45}$ are of orders ${\cal O}(\xi^3\ophi_{45})$ and ${\cal O}(\xi^6)$ respectively. We therefore impose 
\begin{align}
\oPhi_{12} \ophi_{45}\!-\!\overline{B}^{(5)} \Phi_{31}\simlt \xi^6\, ,
\end{align}
so that $\oH_{45}$ is the dominant component in the $\oH$ Higgs doublet. This can be done for instance by choosing $\ophi_{45}\sim\xi^3$ and tuning slightly the combination $\left(\oPhi_{12}-\{D_1^2+D_4^2+T_1^2\}\right)$.
%{\tt the next 3 lines have to be updated}\\
As a result, we obtain:
\begin{align}\label{Hd1}
H&=\cos\theta\, H_1-\sin\theta\, H_{45}\quad;\quad \tan\theta=\langle\oPhi_{31}\rangle/\langle\phi_{45}\rangle\\
%\oH&\simeq\cos\bar\theta\,\oH_{45}+\sin\bar\theta\,\oH_2\quad;\quad \tan\bar\theta=
%\frac{A^{(7)}\langle\Phi_{31}\rangle\!-\! \langle\oPhi_{12}\rangle\langle\oPhi_{23}\rangle}
%{\langle\oPhi_{12}\rangle \langle\ophi_{45}\rangle\!-\!\overline{B}^{(5)} \langle\Phi_{31}\rangle}
%\oH&=\bar c_{45}\oH_{45}+\bar c_2\oH_2+\bar c_1\oH_1\quad;\quad\bar c_{45}^2+\bar c_2^2+\bar c_1^2=1\\
\oH&=\cos\bar\theta\,\oH_{45}+\sin\bar\theta\cos\bar\vartheta\,\oH_2+\sin\bar\theta\sin\bar\vartheta\,\oH_1\quad;\\
% &{\rm with} \quad \frac{\bar c_2}{\bar c_{45}}=
&\tan\bar\theta\cos\bar\vartheta=
\frac{\langle\oPhi_{12}\rangle \langle\ophi_{45}\rangle\!-\!\overline{B}^{(5)} \langle\Phi_{31}\rangle}
{A^{(7)}\langle\Phi_{31}\rangle\!-\! \langle\oPhi_{12}\rangle\langle\oPhi_{23}\rangle}
\quad;\quad %\frac{\bar c_1}{\bar c_{45}}=
\tan\bar\vartheta=
\frac{\overline{B}^{(5)}\langle\oPhi_{23}\rangle\!-\!A^{(7)} \langle\ophi_{45}\rangle}
{\langle\oPhi_{12}\rangle \langle\ophi_{45}\rangle\!-\!\overline{B}^{(5)} \langle\Phi_{31}\rangle}
%{A^{(7)}\langle\Phi_{31}\rangle\!-\! \langle\oPhi_{12}\rangle\langle\oPhi_{23}\rangle}
\,.\nonumber
\label{Hu}
\end{align}
It follows that 
\begin{align}
H_1&=\cos\theta H +\dots\,\,;\,\, H_{45}=\sin\theta H+\dots \nonumber\\
\oH_{45}&=\cos\bar\theta\,\oH+\dots\,\,;\,\, \oH_2=\sin\bar\theta\cos\bar\vartheta\,\oH+\dots
\,\,;\,\,\oH_1=\sin\bar\theta\sin\bar\vartheta\,\oH+\dots
\end{align}
where the dots stand for (superheavy) massive doublets. To simplify the analysis, in the following we will take $\cos\theta\sim\cos\bar\theta\sim 1$.

\subsection{Fermion masses}
We notice that the physical Higgs which provides masses to the charged $Q=2/3$ up quarks is a mixture of $\oH_{45}$ (as a necessary leading component), $\oH_2$ and $\oH_1$, while for $Q=-1/3$ down quarks and charged leptons, the physical Higgs is a mixture of $H_1$ (as a necessary leading component) and $H_{45}$. %It is indeed interesting that 
Indeed, the leading electroweak Higgs components are exactly what the trilinear superpotential~\eqref{wtree} suggests in order to provide masses to the heaviest third generation: $t$, $b$ and $\tau$: %are very succesfull.
\be
W_3\supset  g_s\sqrt{2} \left\{ 
%\sin\bar\omega F_4 \of^{(0)}_5 \oH_{45} + (F_4 F_4 + \of_1\ell^c_1) H_1\right\}\,,
F_4 \of_5 \oH_{45} + (F_4 F_4 + \of_1\ell^c_1) H_1\right\} \,.
\ee
%where $\tan\bar\omega=\langle\ophi_2\rangle/\left(\langle T_3\rangle\!\cdot\!\langle T_4\rangle\right)$ and $\of^{(0)}_5$ is the massless combination orthogonal to \eqref{fbarlc34} containing the right-handed top quark.
Actually, more than thirty years ago, this model predicted that the mass of the top quark is around $\sim$ 170-180 GeV~\cite{Antoniadis:1989zy}, as was observed in 1995 in Fermilab. This is a consequence of the fact that the top Yukawa coupling %is given by the gauge coupling at the grand unification (GUT) scale and 
evolves at low energies towards a fixed point for its ratio to the QCD gauge coupling.
Furthermore, we get the relation at the GUT scale  %λ-τ=λ-b.
$m_b=m_\tau$, following the equality of the corresponding Yukawa couplings, which is apparent 
%Thus it becomes apparent that 
from the above expression of the trilinear superpotential. This is a successful mass relation assuming 
a suitable supersymmetric spectrum \cite{ParticleDataGroup:2020ssz, cite114}.
%Here $F_4$ contains the $(t,b)_L$ left-handed doublet, $b^c_L$ and $\nu^c_\tau$, while $fbar_5$ contains $t_L^c$ and $\of_1$ contains the $\tau$ lepton doublet.
Hence, we have the particle identification for the third generation:
\be
\label{3rdGen}
F_4=\left\{ (t,b), b^c, \nu^c_\tau \right\}_L \quad;\quad %\of^{(0)}_5\supset t^c_L  
\of_5\supset t^c_L\quad;\quad
\of_1\supset (\nu_\tau, \tau)_L \quad;\quad l^c_1=\tau^c_L
\ee

We next look for possible fermion mass terms at higher NR orders in the superpotential involving the massless Higgs doublets \eqref{Hd} and \eqref{Hu0}. Omitting corrections to the Yukawa couplings of the 3rd generation, we find the following list up to 5th order:
%\noindent
%List of candidate fermion mass terms:\\
\begin{itemize}
\item
Up quarks:
\begin{gather}
%F_3 \of_3 \oh_3 + 
%F_4 \of_5 \oh_{45}\\
%F_1 \of_1 \oh_{45} \phi_1+
F_2 \of_2 \oh_{45}\ophi_4 \label{charm}
%F_1 \of_1 \oh_{45} \phi_1 \Phi_2 + F_1 \of_5 \oh_{45} (T_1 T_4) +
%F_2 \of_2 \oh_{45} \ophi_4 \Phi_1 %+ F_3 \of_3 \oh_1 D_5^2 + 
%F_3 \of_3 \oh_2 T_4^2 %+ F_4 \of_2 \oh_{45} (D_2 D_5)
\end{gather}
\item
Down quarks:
\begin{gather}
%F_1 F_1 h_1 + F_2 F_2 h_2 + 
%F_4 F_4 h_1\\
%F_1 F_1 h_2 \left(\phi_1^2+\phi_2^2+\phi_3^2+\phi_4^2+\phi_+ \phi_-\right)
%+ F_1 F_1 h_3 D_5^2 + F_1 F_1 h_{45} \phi_{45} \Phi_{31}\\
F_2 F_2 h_1 \left(\ophi_1^2%+\ophi_2^2+\ophi_3^2
-\ophi_+ \ophi_- +\lambda %(
\ophi_4^2 %+\ophi_2^2)
\right) \label{squark}
%+ F_2 F_2 h_3 T_4^2 
%+F_2 F_2 h_{45} \phi_{45} \oPhi_{23}
%\\
%+ F_4 F_4 h_2 \left(\phi_1^2+\phi_2^2+\phi_3^2+\phi_4^2+\phi_+ \phi_-\right)
%+ F_4 F_4 h_3 D_5^2 
%+ F_4 F_4 h_{45} \phi_{45} \Phi_{31}
\end{gather}
\item
Charged leptons: %(except for right-handed neutrinos):
\begin{gather}\label{cl}
%\of_1 \ell^c_1 h_1 %+ \of_2 \ell^c_2 h_2 +  \of_5 \ell^c_5 h_2 \\
%+ \of_1 \ell^c_1 h_2 \left(\phi_1^2+\phi_2^2+\phi_3^2+\phi_4^2+\phi_+ \phi_-\right) + \of_1 \ell^c_1 h_3 D_5^2
 %+ \of_1 \ell^c_1 h_{45} \phi_{45} \Phi_{31}
% \\%+ 
\of_2 \ell^c_2 h_1 \left(\ophi_1^2%+\ophi_2^2+\ophi_3^2
-\ophi_+ \ophi_- +\lambda %(
\ophi_4^2 %+\ophi_2^2)
\right) 
%+ \of_2 \ell^c_2 h_3 T_4^2 
%+ \of_2 \ell^c_2 h_{45} \phi_{45} \oPhi_{23}\\
+\of_5 \ell^c_5 h_1 \left(\ophi_1^2%+\ophi_2^2+\ophi_3^2
+\ophi_+ \ophi_- -\ophi_4^2
\right) 
%+ \of_5 \ell^c_5 h_3 T_4^2 
%+ \of_5 \ell^c_5 h_{45} \phi_{45} \oPhi_{23}\,,
\end{gather}
\end{itemize}
where we displayed only the non-vanishing dominant contributions for the choice of VEVs that solved the flatness conditions in Table~\ref{tableofvevs},
and we made a convenient choice for the co-cycle factors ambiguity in the 5-point amplitudes fixing the relative signs, while %. We also present all relative signs, with 
$\lambda$ is an (irrelevant) order one constant %(since $\ophi_4\sim\xi^2$).
(see below).

Using the flatness condition \eqref{fPhib125}:
\be
\ophi_1^2 +\ophi_+ \ophi_- + \ophi_4^2 %+\ophi_2^2 
= -\oPhi_{31}\oPhi_{23} = {\cal O}(\xi^4)\, ,
\label{phibari2}
\ee
%By inspection of the above couplings and using that $\langle\ophi_4\rangle\sim{\cal O}(\xi^2)$ while the rest of the VEVs are of order ${\cal O}(\xi)$, 
one can identify all members of the 2nd generation: %with Yukawa couplings of the same order:
\be
\label{2ndGen}
F_2=\left\{ (c,s), s^c, \nu^c_\mu \right\}_L \quad;\quad \of_2= \left\{ c^c, (\nu_\mu,\mu) \right\}_L
 \quad;\quad l^c_2=\mu^c_L\,,
\ee
%Note that the 
with Yukawa couplings %of the 2nd generation are s
suppressed by %roughly 
two orders of magnitude compared to those of the 3rd generation. 
Indeed, \eqref{charm} provides a successful mass to the charm quark when $\ophi_4\sim{\cal O}(\xi^2)$, while \eqref{squark} and \eqref{cl} provide successful masses to the strange quark and muon.
Moreover, a direct computation of the corresponding coefficients of the above 5th order operators shows that those of $F_2F_2h_1$ are equal to those of $\of_2l^c_2h_1$~\cite{Lopez:1991ac}. One thus obtains the mass relation $m_s=m_\mu$.

It remains the identification of the first generation. Taking into account the mass term \eqref{fbarlc34}, the identification \eqref{3rdGen}, \eqref{2ndGen} and the leftover operator for the charged leptons Yukawa couplings, one gets:
\be
\label{1stGen}
F_3=\left\{ (u,d), d^c, \nu^c_e \right\}_L \quad;\quad \of_5%^{(0)}
\supset (\nu_e, e)_L  \quad;\quad
\of_1\supset u^c_L \quad;\quad l_5%^{c(0)}
=e^c_L\,.
\ee
Combining \eqref{3rdGen} and \eqref{1stGen}, we thus obtain:
\be
\label{3rd1stMix}
\of_5%^{(0)}
=\left\{ t^c, (\nu_e, e) \right\}_L \quad;\quad \of_1=\left\{ u^c, (\nu_\tau, \tau) \right\}_L\,.
\ee
%Then using the NR higher ( than 3) orders ,as given in eqs(6.7) through (6-17) we see that ...F_2,f_2bar,l_2^c should be identified with the second generation, While F_3 contains the (u,d)_L ,d_L^c,while fbar_1 contains the u_L^c and fbar_5 Contains the electron doubled,while the e_L^c is identified with l_5^c. 
It is amazing how in this string model the three generations are distributed with rather interesting consequences for inter-generation mixing, proton decay modes and possibly flavour changing lepton number.
%FCLepton N... Using eqs...( here I mean LN eq 2,3,4) , and our vev solutions of the flatness eq See sections 4,5, we find: .....and we put here all we get:
%i)t,b,τ   Ii) λ_c ,λ_s,λ_μ Ιιι)λ_e and .λ_d...
%As absolute values BUTalso we get some relations Between masses ,for the first time,that are successfull And INDEPENDENT Of the specific vevs we are using,,,,,,,,!!!!!!!!
Up to this order, there are no Yukawa coupling for the up and down quarks, which is compatible with the order of our approximation. On the other hand, the second operator in \eqref{cl} leads to a Yukawa coupling for the electron. Using the flatness relation \eqref{phibari2} and the identification \eqref{3rd1stMix}, one finds that 
%the electron Yukawa coupling gets a $\sin\bar\omega^2$ suppression and thus 
%It turns out that the coefficients of $\ophi_1^2$ and $\ophi_+\ophi_-$ are equal and using the tree-level relation \eqref{phibar1+-} required by the flatness conditions, one finds that 
the electron mass is two orders of magnitude lower than the muon mass, which is a remarkable successful relation~\cite{Lopez:1991ac}.

Focussing now to possible mass terms for the quarks of the first generation, we extracted all relevant NR terms up to 7th order (included) for our choice of VEVs, that we display below, omitting higher order corrections to existing Yukawa couplings:
\begin{itemize}
%\noindent
%N=6 mass terms : \\
\item
Up quarks: %\\
%\begin{align}
\begin{gather}
F_1 \of_1 \oh_2 \Phi_{31}\phi_{45}\ophi_1+
%F_1 \of_1\oh_3 \oPhi_{12} \ophi_1 \phi_{45} + F_2 \of_2 \oh_1 \oPhi_{23} \phi_{45} \phi_4 \\
%+ F_2 \of_2 \oh_{45} \left(\Phi_{23}\oPhi_{23} +\Phi_{31}\oPhi_{31}\right)\ophi_4 + 
{ F_3 \of_5 \oh_{45} (D_3 D_4) \oPhi_{23}} \nonumber\\
+F_3 \of_1 \oh_{45}  \left[\,\ophi_-\left(T_1 T_3\right)\Phi_{31}+\ophi_1 \left(D_1 D_3\right)\Phi_{31}\right]
\label{uq2}
\end{gather}
%\end{align}
\item
Down quarks: %\\
%\begin{align}
\begin{gather}\label{dq2}
F_1 F_3 h_1 \left(D_1 D_3\right)\oPhi_{23}
%+F_1 F_3 h_2 \left(D_1 D_3\right) \Phi_{31} 
+F_3 F_4 h_1 \left(D_3 D_4\right)\oPhi_{23} 
%+F_3 F_4 h_2 \left(D_3 D_4\right) \Phi_{31}
%\end{align}
\end{gather}
\item
Charged leptons: %\\
%\begin{align}
\begin{gather}\label{cl2}
%\quad \of_1 \ell^c_3 h_1 \left(T_1 T_3\right)\oPhi_{23} +
%%\of_1 \ell^c_3 h_2 \left(T_1 T_3\right)\Phi_{31} +
%\of_3 \ell^c_1 h_1 \left(T_1 T_3\right)\oPhi_{23} 
%%+ \of_3 \ell^c_1 h_2 \left(T_1 T_3\right)\Phi_{31}
\of_5 \ell^c_5 \left( h_1 \oPhi_{31} +h_{45}\phi_{45}\right) \left(\ophi_1^2+\ophi_4^2+\ophi_+\ophi_-\right)\Phi_{31}
%\of_5 \ell^c_5   \left(\ophi_1^2+\ophi_4^2+\ophi_+\ophi_-\right)\Phi_{31}
\end{gather}
%\end{align}
\end{itemize}
Eq.~\eqref{cl2} gives a %contribution
correction to the electron Yukawa coupling of order ${\cal O}(\xi^6)$. %${\cal O}(\xi^4)$, comparable to the contribution \eqref{cl} due to the cancellation we described above. 
The term in the second line of \eqref{uq2} leads to an up quark Yukawa coupling of the right order of magnitude ${\cal O}(\xi^5)$, provided $T_3$ and $D_3$ are of order ${\cal O}(\xi^2)$, consistently with the 6th order flatness conditions discussed at the end of section~\ref{FDsol}. Finally, the second term in the first line of \eqref{uq2}, as well as the second term of \eqref{dq2} lead to quark mixing between the first and third generation of order ${\cal O}(\xi^6)$. %and ${\cal O}(\xi^5)$, respectively.

%Selected N=7 mass terms \\
%Up quark:
%\begin{align}
%F_3 \of_1 \oh_{45}  (\ophi_-\left(T_1 T_3\right)\Phi_{31}+\ophi_1 \left(D_1 D_3\right)\Phi_{31})
%%+F_4 \of_5 \oh_1 \left(\phi_{45}\ophi_{45} + \phi_+\ophi_++\phi_-\ophi_-+\phi_4\ophi_4\right)%\phi_{45}\Phi_{31} \\
%%F_4 \of_5 \oh_1 \left(\phi_+\ophi_++\phi_-\ophi_-+\dots\right)\phi_{45}\Phi_{31}
%%+F_4 \of_5 \oh_{45} \left(\phi_-^2\ophi_-^2+\phi_+^2\ophi_+^2+\dots\right)
%\end{align}
%Down quark: \\
%\begin{align}
%%F_2 F_2 h_1 \left(\ophi_1^2+\ophi_4^2+\ophi_+\ophi_-\right)\Phi_{31}\oPhi_{31}+
%%F_2 F_2 h_{45} \left(\ophi_1^2+\ophi_4^2+\ophi_+\ophi_-\right) \Phi_{31} \phi_{45}+\\
%%F_4 F_4 h_1 \left(\phi_-^2\ophi_-^2+\phi_+^2\ophi_+^2+\dots\right)+
%%F_4 F_4 h_{45} \phi_{45} \Phi_{31}^2\oPhi_{31}
%%\end{align}
%Lepton:\\
%\begin{align}
%%\of_1 \ell^c_1 h_1 \left(\phi_-^2\ophi_-^2+\phi_+^2\ophi_+^2+\dots\right)+
%%\of_1 \ell^c_1 h_{45} \phi_{45} \Phi_{31}^2\oPhi_{31}+\\
%%\of_2 \ell^c_2 h_1  \left(\ophi_1^2+\ophi_4^2+\ophi_+\ophi_-\right)\Phi_{31}\oPhi_{31}+
%%\of_2 \ell^c_2 h_{45} \phi_{45} \left(\ophi_1^2+\ophi_4^2+\ophi_+\ophi_-\right)\Phi_{31}+\\
%\of_5 \ell^c_5 h_1  \left(\ophi_1^2+\ophi_4^2+\ophi_+\ophi_-\right)\Phi_{31}\oPhi_{31}+
%\of_5 \ell^c_5 h_{45}  \left(\ophi_1^2+\ophi_4^2+\ophi_+\ophi_-\right)\Phi_{31}\phi_{45}
%\end{align}

A down quark Yukawa coupling can appear when $F_3$ gets a small VEV which is consistent with the flatness conditions when $\langle F_3\rangle/\langle F_1\rangle<\xi$, as explained in the end of section 4. In this case, the GUT Higgs $F$ and the first generation $F_3'$ are given by the linear combinations:
\bea
F &=& A_1F_1+A_3F_3 \simeq F_1+\varepsilon F_3 \nonumber\\
F_3' &=& -A_3F_1+A_1F_3 \simeq F_3 -\varepsilon F_1
\label{fmm}
\eea
where the constants $A_i$ satisfy $\sum_i |A_i|^2=1$ and enter in the flatness condition \eqref{FFbar}, while $\varepsilon\sim{\cal O}(\xi^{3/2})$. Thus, the tree-level superpotential term $F_1F_1h_1$ in \eqref{wtree} generate a Yukawa coupling for the down quark of the right order of magnitude ${\cal O}(\xi^3)$. 

Our results on the masses of quarks and leptons and their relations (at the string scale) are summarised below:
\be
\label{fermionmasses}
\begin{matrix}
m_t=gv_{45}  & m_b=gv_1 & m_\tau=m_b \\
m_c\sim\xi^2 m_t & m_s\sim\xi^2 m_b & m_\mu=m_s \\
m_u\sim\xi^5 m_t & m_d\sim \xi^3 m_b & m_e\sim \xi^4 m_\tau \\
\end{matrix}
\ee
where $g=g_s\sqrt{2}$ is the GUT gauge coupling and $v_{45}, v_1$ denote the 
VEVs of $\oH_{45}, H_1$ Higgs doublets, respectively (see Eqs. \eqref{Hu}, 
\eqref{Hd1}). It follows that $\tan\beta\sim m_t/m_b\simeq 40$. 

The relation $m_\mu=m_s$ is apparently problematic. Moreover, we have
not obtained a $d-s$ quark mixing. Both issues can be in principle addressed by allowing an appropriate non-zero VEV 
for $F_2$ generalising the flipped $SU(5)$ breaking mixing VEVs in \eqref{fmm}. This introduces a mixing between $F_2$ 
and $F_3$ which could account for the Cabibo angle and correct the relation $m_\mu=m_s$. However,  a separate analysis 
is needed that could also include neutrino masses and mixings which goes beyond the scope of this paper.

We consider that equation \eqref{fermionmasses} belongs to the highlights of this work, and as such, we need to pause 
and reflect on its importance. We have achieved, for the first time ever to our knowledge, to compute explicitly the 
mass spectrum of quarks and (charged) leptons of the Standard Model in String Theory. By calculating the superpotential 
at the $N=3$ (tree) level, we identified the content of the third generation, i.e. the particles that get Yukawa 
couplings proportional to the string coupling constant $g_s$ at this level. We discussed above about the top quark 
Yukawa coupling and its consequences of predicting in 1989 the top quark mass in the 170-180 GeV range, as observed in 
1995 at FNAL to be around 173 GeV. This particular top-quark Yukawa coupling triggers the radiative electroweak 
breaking of $SU(2)\times U(1)$ at low energies, thus explaining the gauge hierarchy $M_W/M_{\rm GUT}\sim{\cal 
O}(10^{-16})$ a natural way. Actually, because our string model is of no-scale type, it leads to a determination of the 
SUSY breaking scale in the ${\cal O}({\rm TeV})$ region. Concerning the masses of the bottom quark and $\tau$-lepton, 
we get the relation $m_b=m_\tau$ at the string scale, as well as the equality of the top and bottom Yukawa couplings 
that leads to the determination of $\tan\beta\sim m_t/m_b\simeq 40$, which eventually would be determined dynamically 
through the no-scale mechanism. 

For the next two generations, we need to calculate non renormalisable corrections in $\alpha'$ in the superpotential, corresponding to $N=4, 5, 6,\dots$, that we have done using the general method of Ref.\cite{Kalara:1990fb}. The $N$-th order NR terms contain $(N-3)$ fields that will need to get VEVs. The way that all these fields get their VEVs is through the endemic, in the string models we are considering, existence of an `anomalous' abelian  gauge symmetry $U(1)_A$ that enforces non-trivial VEVs for some `charged' fields. Eventually, in order to satisfy the F- and D-type flatness conditions, a set of fields get dynamically VEVs of the order $\xi\approx 1/10 M_s$, and thus we have a perturbative expansion parameter! Thus, all the masses of the second and first generations are found to be determined, involving powers of $\xi^n$, $n=2,3,4,5$, multiplying, for normalisation, the corresponding masses of the third generation. In other words, the masses of the second and first generations are dynamically determined as $m_t, m_b, m_\tau$ and $\xi$ and are dynamically fixed! 

Now, we can really appreciate the structure of Eq.~\eqref{fermionmasses}, as it provides a very successful mass estimation for all quarks and charged leptons.

\subsection{Proton decay}
Let us now turn to the problem of proton decay. Proton decay 
has been for more than forty years a real headache for theorists. Basically,
it is a main prediction of GUTs that has not been vindicated experimentally. 
The present lower limits on proton decay are of the order of $10^{34}-10^{35}$ 
years depending on the particular decay mode. Unlike the Standard Model 
where one can show that, because of its particle content, 
contains no baryon (B) and lepton (L) number violating interactions \cite{PD} in Grand 
Unified Theories, these interactions are endemic.
% They led to the demise of 
%non-supersymmetric GUTs, while supersymmetric (SUSY) GUTs need enforcements to avoid 
%rapid proton decay, including matter-parity.
Furthermore, SUSY GUTs contain dangerous $d=5$ B and L violating interactions 
that may lead to very rapid proton decay. With the advent of superstring 
theory, generally, the proton decay problem became more acute. The reason being 
that the low energy spectrum contains a plethora of particles that may provide 
B, L violating interactions leading to a rather rapid proton decay.

% Even if 
%this problem is resolved, there are effective non-renormalisable interactions 
%mediated by massive string states with an amplitude of the order of 
%$M_\text{string}^{-1}$ that is not enough to avoid rapid proton decay. 
%Though, all hope is not lost. There is an exception, if we use the Flipped 
%$SU(5)$  or $SU(5)\times{U(1)}$ model (FSU5). 
In the case of 
%the supersymmetric GUTs the dominant nucleon 
%decay mode is mediated by dimension five operators.   
%\cite{Ellis:2020qad}
%For the 
the string derived flipped $SU(5)$ model under consideration there are two sources 
of $d=5$ baryon number violating operators. The first consists of the usual dimension five operator 
$QQQL$ ascribed to the exchange of additional triplets in the massless string 
spectrum. The second comprises effective $QQQL$ operators generated from 
non-renormalisable string couplings arising from the exchange of massive string 
modes. 

Let us start with the triplet exchange induced dimension-five operators. As explained in Section 3 we have five pairs of additional triplets accommodated in the fields $h_i, \oh_i, i=1,2,3,45$ and the flipped $SU(5)$ breaking Higgs multiplets $F, \oF$ defined in this section. At tree-level the triplets mass matrix is 
given by \eqref{tripmm} where $d^c_H$ now stands for the additional triplet combination $d^c_1+\varepsilon d^c_3$ and $F_1$ is replaced by $F$.
For our flatness solution of Table \ref{tableofvevs} and taking into account non-renormalisable interactions up to $N=7$, the extra triplet mass matrix  
to order $\xi^5$ reads
\begin{align}
	M_D^{(7)} = \bordermatrix{
		~&D_1&D_2&D_3&D_{45}&\overline{d}^c_H\cr
		\bar{D}_1&0&\oPhi_{12}&\Phi_{31}&0&\overline{s}_1^{(5)}\cr
		\bar{D}_2&0&A^{(7)}&\oPhi_{23}&0&2\oF_5\cr
		\bar{D}_3&\oPhi_{31}&\Phi_{23}&0&\phi_{45}&\overline{s}_3^{(5)}\cr
		\bar{D}_{45}&0&\overline{B}^{(5)}&\ophi_{45}&0&0\cr
		d^c_H&2F&s_2^{(5)}&0&s_4^{(5)}&s\cr
	}+{\cal O}\left(\xi\right)^6\,,
		\label{ftripmm}
\end{align}
where
\begin{align}
\overline{s}_1^{(5)} &= 
\left\{\oF_5\left(\phi_4^2+\phi_+\phi_-\right)\right\}\,,\\
\overline{s}_3^{(5)} &=  \left\{\oF_5\left(D_1^2+D_4^2+T_1^2\right)\right\}\,,\\
{s}_2^{(5)} &=  \left\{F \left(\phi_4^2+\phi_+\phi_-\right)\right\}\,,\\
{s}_4^{(5)} &=  \left\{F \Phi_{31} \phi_{45}\right\}\,,\\
s&= \left\{F \oF_5 \oPhi_{12}\right\}\,.
\end{align}
Assuming $F \oF_5 \sim \xi^3$ the determinant is given by
\begin{align}
\det\left(M_d^{(7)}\right) &\sim 
\left( F \phi_{45}- \oPhi_{31} {s}_4^{(5)}\right)
\left(
\overline{B}^{(5)} \oF_5\Phi_{31}- \oF_5 \oPhi_{12}\ophi_{45} 
-\overline{B}^{(5)} \overline{s}_1^{(5)} \oPhi_{23} +
A_2^{(7)} {\overline{s}}_1^{(5)} \ophi_{45}
\right)\nonumber\\
&\sim F \oF_5    \phi_{45} \left(\overline{B}^{(5)} \Phi_{31} 
-\oPhi_{12}\ophi_{45}\right)+ \dots \sim \xi^{10}
\end{align}
ensuring that all triplets are massive. A detailed calculation shows that the orders of magnitude of the triplets mass eigenstates are: 
$\xi, \xi, \xi^\frac{3}{2}, \xi^\frac{3}{2}, \xi^5$. This is consistent with 
our approximation utilised in \eqref{ftripmm}, as the lightest 
eigenvalue is of order ${\cal O}(\xi^5)$ rendering higher order contributions in \eqref{ftripmm} irrelevant. 

Following the analysis of \cite{Gomez:1998zf}, triplet exchange $QQQL$ type 
dimension-five operators for a 
general superpotential  of the form 
\begin{align}
f_{ij}^a  F_i F_j h_a + y_{ij}^a F_i \bar{f}_j \bar{h}_a + \oh_a \left(M_D\right)_{ab} h_b
\end{align}
involving $n$ additional fiveplets are proportional to 
\begin{align}
{\cal O}^{QQQL}_{ijkl} \sim \frac{1}{\det\left(M_D^{(7)}\right)} \sum_{a,b=1}^n y_{kl}^a
\,{\rm cof}\left(M_D^{(7)}\right)_{ab} f_{ij}^b \,.
\label{qqqle}
\end{align}
In our case, a direct computation yields
\begin{align}
{\rm cof}\left(M_D^{(7)}\right)_{1,1} &= \oF_5 \phi_{45} \ophi_{45} s_2^{(5)} + 
\dots \sim \xi^9
\label{cfi}\,,\\
{\rm cof}\left(M_D^{(7)}\right)_{1,2} &= - \phi_{45} \ophi_{45} s_2^{(5)} 
\overline{s}_1^{(5)} + \dots \sim \xi^{11}\,,\\
{\rm cof}\left(M_D^{(7)}\right)_{1,3} &= \oF_5 s_4^{(5)}\left(\Phi_{12} 
\phi_{45}-\overline{B}^{(5)}\Phi_{31}\right) + \dots \sim \xi^{11}\,,\\
{\rm cof}\left(M_D^{(7)}\right)_{1,45} &= -\oF_5 \Phi_{31} \phi_{45} s_2^{(5)} 
+ \dots \sim \xi^{7}\,,\\
{\rm cof}\left(M_D^{(7)}\right)_{2,1} &= -F_1 \oF^5 \phi_{45} \ophi_{45} 
s_2^{(5)} + \dots \sim \xi^7\,,\\
{\rm cof}\left(M_D^{(7)}\right)_{2,2} &= F_1  \phi_{45} \ophi_{45} 
\overline{s}_1^{(5)} + \dots \sim \xi^9\,,\\
{\rm cof}\left(M_D^{(7)}\right)_{2,3} &= 0\,,\\
{\rm cof}\left(M_D^{(7)}\right)_{2,45} &= F_1 \oF_5 \Phi_{31} \phi_{45}  + 
\dots \sim \xi^{5}\,,\\
{\rm cof}\left(M_D^{(7)}\right)_{3,1} &= F_1 \oF_5 \phi_{45} \overline{B}^{(5)} 
+ \dots \sim \xi^9\,,\\
{\rm cof}\left(M_D^{(7)}\right)_{3,2} &= - F_1 \phi_{45} \overline{B}^{(5)} 
\overline{s}_1^{(5)} + \dots \sim \xi^{11}\,,\\
{\rm cof}\left(M_D^{(7)}\right)_{3,3} &= 0\,,\\
{\rm cof}\left(M_D^{(7)}\right)_{3,45} &= -F_1 \oF_5 \phi_{45} \oPhi_{12} + 
\dots \sim \xi^{7}\label{f3f3d5}\\
{\rm cof}\left(M_D^{(7)}\right)_{45,1} &= -\oF_5 \ophi_{45}\left( \Phi_{31} 
s_2^{(5)}- F_1 \Phi_{23}\right) + \dots \sim \xi^9\,,\\
{\rm cof}\left(M_D^{(7)}\right)_{45,2} &= \left(\oPhi_{31}  s_2^{(5)}   - F_1 
\Phi_{23}\right) \ophi_{45} \overline{s}_1^{(5)} 
\!+\! \left( \oPhi_{12} \ophi_{45} - \Phi_{31} \overline{B}^{(5)} \right) F_1 
\overline{s}_3^{(5)}
\!+\! \dots \sim \xi^{11}\,,\\
{\rm cof}\left(M_D^{(7)}\right)_{45,3} &= -F_1 \oF_5 \left(\oPhi_{12} 
\ophi_{45}-\overline{B}^{(5)} \Phi_{31}\right) + \dots \sim \xi^{9}\,,\\
{\rm cof}\left(M_D^{(7)}\right)_{45,45} &= \oF_5 \Phi_{31} \left(\oPhi_{31} 
s_2^{(5)}-F_1 \Phi_{23} \right) + \dots \sim \xi^{7}\,.
\label{llast}
\end{align}

Triplet-exchange dimension-five operators can be generated via couplings of the 
type $F F h_2 \varphi^{N-3}$ and $\oF\,\of\,\oh_{45} {\varphi'}^{M-3}$, arising 
at orders $N$ and $M$ respectively, where $\varphi^{N-3}$, ${\varphi'}^{M-3}$ 
stand for combinations of field VEVs. At tree-level ($N=M=3$) we have a single 
pair of couplings of this type, namely $F_2 F_2 h_2, F_4 \of_5 \oh_{45}$, that 
can give rise to an effective $F_2 F_2 F_4 \of_5$ operator. Actually, this 
operator corresponds to the dominant 
contribution in the $\xi$ expansion, among \eqref{cfi}-\eqref{llast}, arising  
from ${\rm 
cof}\left(M_D^{(7)}\right)_{2,45}$ which is proportional to
\begin{align}
\frac{{\rm cof}\left(M_D^{(7)}\right)_{2,45}}{\det\left(M_d^{(7)}\right)} \sim 
\frac{1}{\xi^5}\,.
\label{cll}
\end{align}
 However, the associated $d=5$ operator is further suppressed by at least a factor $\xi^3$, 
 since $F_4, F_2$ accommodate third and second generation quarks respectively. 
 In fact the associated $QQQL$ operator is suppressed by an effective triplet 
 mass of order $\xi^2 M_P\sim 10^{16}$ GeV which leads to proton lifetime 
 exceeding current experimental limits for  a SUSY breaking 
 scale of the order of $m_{\rm susy}\gtrsim 10^2$ TeV \cite{Hisano:2013exa}. 
Additional tree-level operators could be generated via the $F_1-F_3$ mixing 
introduced in Section 6.1 to generate down quark mass (see \eqref{fmm}). The 
term $F_1 F_1 h_1$ induces an effective coupling of the form $F_3 F_3 h_1$ that 
combined with the $\oF\,\of\,\oh_{45}$ leads to an effective $d=5$ operator of 
the form  $F_3 F_3 F_4 \of_5$. However, this operator gets an extra 
suppression $\xi^3$ due to the mixing and $\xi^7$ due to Eq.~\eqref{f3f3d5}, 
leading to an effective triplet scale of the order of $M_P$ and thus becomes  
subdominant.

At higher order ($N=3, M=4$)
one could use the terms $F_2 F_2 h_2$ and $F_2 \of_2\oh_{45} \ophi_4$ to form an effective dimension five operator of the type $F_2 F_2 F_2 \of_2$. In this case we have smaller family mixing, of the order of $\xi^2$, however, we get an additional suppression of  order  $\xi^2$ from the additional VEV $\ophi_4$. Furthermore, higher order contributions ($N,M >3$) are relatively suppressed by a factor of $\xi^2$ in the worst case scenario. The same is true for all other operators in Eqs. \eqref{cfi}-\eqref{cll}. 

Let us now examine the string induced effective dimension-five operators \cite{Ellis:1990vy}. These are of the form ${\mathbf{10}\!\times\!\mathbf{10}\!\times\!\mathbf{10}}\times{\overline{\mathbf{5}}}$. An explicit search 
gives no candidate couplings of this type at the level of $N=4$ 
non-renormalisable superpotential, while at $N=5$ we have the following two 
terms
\begin{align}
F_2^2 F_3 \of_3 \oPhi_{23} + F_4^2 F_3 \of_3 \Phi_{31}\,.
\label{opta}
\end{align}
However, both terms involve $\of_3$ which becomes superheavy in our F/D-flatness solution (see \eqref{fbarlc34}). As a result, we have no contributions to proton decay at this level. At $N=6$ we find two non-vanishing terms
\begin{align}
F_3^2 F_4 \of_5 \left(\Phi_{31} \ophi_- + \oPhi_{23} \phi_+\right)\,.
\label{optb}
\end{align}
 Following  Table \ref{tableofvevs}, these  yield an affective operator $F_3^2 F_4 \of_5$ with a coupling of order $\xi^2$ which is translated to 
a dimension-five $QQQL$ effective operator with triplet scale of the order of 
$\xi^{-2} M_P \sim 10^{20}$ GeV, which as explained above is safe for proton 
decay. Moreover, all operators of the form \eqref{opta} have been shown to 
vanish explicitly in the case of the flipped $SU(5)$ model as a result of 
permutation symmetries \cite{Ellis:1990vy}.

Summarising, the leading contribution to dimension-five proton decay operators comes from the $F_2 F_2 F_4 \of_5$ 
operator arising from $h_2, \oh_{45}$ triplet pair mediation which is compatible with the experimental bounds for a 
SUSY breaking scale $m_{\rm susy}\gtrsim {\cal O}(10^2)$ TeV. It is interesting to point out that a SUSY breaking scale 
in the energy region of tens of TeV is also required for cosmological reasons, following an analysis of reheating and 
nucleosynthesis in the flipped $SU(5)\times U(1)$ model~\cite{Ellis:2019opr}. The model can also accommodate the usual  
lightest supersymmetric particle (LSP) as a sufficiently stable dark matter candidate~\cite{Ellis:1983ew}.

\section{Concluding remarks}

The quest for a Unified Theory of all interactions, including gravity, has been for the last hundred years the `holly grail' of High Energy physics. In our times, it has been named the Theory of Everything (TOE), and as such it should explain not only all of particle physics but also inflationary cosmology in terms of some fundamental principles. Superstring theory has been heralded as the fundamental framework that has the capacity to provide such a Theory of Everything. There are different formulations of (compactified) superstring theory in four dimensions, and for more than thirty years now, the Free Fermionic Formulation (FFF) has been a very useful tool to perform explicit calculations and construct models that may serve as a TOE. 

Recently, we derived from the FFF of superstring theory, a Starobinsky like inflationary model that fits all known cosmological data and connects the inflation scale, calculated dynamically, to the Right-handed neutrino mass, as the inflaton field is a mixture of the heavy sneutrino and some GUT singlet fields provided by superstring theory. The framework is superstring derived no-scale flipped $SU(5)$, that has some unique features, as we discussed in previous works.

Here, we worked out in great detail all the possible physics issues that needed to be resolved. We proved that the F- and D-flatness conditions are satisfied at least to sixth order in the $\alpha'$-expansion of the superpotential, taking into account the fact that in our framework there is always an `anomalous' abelian gauge symmetry. The breaking mechanism of this anomalous $U(1)_A$ entails several fields, mostly singlets, to get VEVs of order $\xi\approx 1/10$ in string units. We use then  $\xi$ as an expansion parameter in perturbation theory, and thus we solve the F- and D-flatness conditions and get a specific set of VEVs dynamically. Then, we are using this `vacuum' to determine the triplet-doublet Higgs splitting and getting a pair of `massless' Higgs doublets that provides the radiative electroweak symmetry breaking and the Yukawa couplings for the third generation at the tree level of the superpotential. Actually, about 32 years ago, we predicted the mass of the top-quark in the range of 170-180 GeV~\cite{Antoniadis:1989zy}. 

Furthermore, non-renormalisable terms, calculable in our framework, provide a realistic hierarchical fermion mass spectrum with all quark and lepton masses derived dynamically, consistent with the experimental hierarchies. As an example, we mention that for the first time ever, the mass of the electron has been calculated explicitly and in full agreement with its observed value, which is rather remarkable. taking into account the fact of its tiny value, vis a vis the top-quark mass. In addition, we derived some new relations involving quarks and leptons that are experimentally satisfied. Furthermore, the triplet Higgs masses are heavy enough as to provide a possible observable proton decay in very specific modes. 

We believe that, all in all, we have for the first time a framework that provides not only a unified picture of particle physics and cosmology, but also a dynamically derived hierarchical mass spectrum, probably observable proton decay and inflationary cosmology in agreement with all cosmological data. We cannot avoid but close with the same final statement used 32 years ago in the first of reference \cite{Ellis:1990vy}: ``We leave it to the reader to decide how many more miracles she wants to see before abandoning her doubts about flipped $SU(5)$".

%\vspace{-0.5cm}
%\enlargethispage{1cm}
\section*{Acknowledgements}
Work partially performed by I.A. as International Professor of the Francqui Foundation, Belgium. The work of DVN was supported in part by the DOE grant DE-FG02-13ER42020 at Texas A\&M University and in part by the Alexander S. Onassis Public Benefit Foundation.
 DVN  would like to thank his beloved wife Olga and the adorable young 
 Odysseas  for their continuous support, encouragement and inspiration during 
 the long period  that this
work was in progress.

\newpage
\begin{appendices}
\section{String construction and spectrum of the flipped $SU(5)\times{U(1)}$ model}
In this appendix we briefly review the construction and the spectrum of the revamped flipped $SU(5)\times{U(1)}$ string model \cite{Antoniadis:1989zy}.
The model is defined in the free fermionic formulation of the heterotic superstring \cite{Antoniadis:1986rn} using a set of eight basis vectors $B=\left\{\beta_1,\beta_2,\dots,\beta_8\right\}$ and a set of phases $c_{ij}=c\left[\beta_i\atop\beta_j\right]\,,i,j=1,\dots,8$ where
\begin{align}
\beta_1 =S & = \left\{\psi^\mu,\chi^{1,\dots,6}\right\}\,,\nonumber\\
\beta_2 =b_1 & = \left\{\psi^\mu,\chi^{12},y^{3,\dots,6};\overline{y}^{3,\dots,6},\overline{\psi}^{1,\dots,5},\overline{\eta}^1\right\}\,,\nonumber\\
\beta_3 =b_2 & = \left\{\psi^\mu,\chi^{34},y^{12},\omega^{56};\overline{y}^{12},\overline{\omega}^{56},\overline{\psi}^{1,\dots,5}\overline{\eta}^2\right\}\,,\nonumber\\
\beta_4 =b_3 & = \left\{\psi^\mu,\chi^{56},\omega^{1,\dots,4};\overline{\omega}^{1,\dots,4},\overline{\psi}^{1,\dots,5},\overline{\eta}^3\right\}\,,
\label{rfbasis}\\
\beta_5 =b_4 & = \left\{\psi^\mu,\chi^{12},y^{36},\omega^{45};\overline{y}^{36},\overline{\omega}^{45},\overline{\psi}^{1,\dots,5},\overline{\eta}^1\right\}\,,\nonumber\\
\beta_6 =b_5 & = \left\{\psi^\mu,\chi^{34},y^{26},\omega^{15};\overline{y}^{26},\overline{\omega}^{15},\overline{\psi}^{1,\dots,5},\overline{\eta}^2\right\}\,,\nonumber\\
\beta_7 =\zeta & = \left\{\overline{\phi}^{1,\dots,8}\right\}\,,\nonumber\\
\beta_8 =\alpha & = \Bigl\{y^{46},\omega^{46};\overline{y}^{46},\overline{\omega}^{2346},
\underbrace{\overline{\psi}^{1,\dots,5},\overline{\eta}^{1,2,3},\overline{\phi}^{1,\dots,4}}_{\frac{1}{2},\dots,\frac{1}{2}},\overline{\phi}^{56}\Bigr\}\,,\nonumber
\end{align}
and
$c_{ij} = c\left[\beta_i\atop\beta_j\right]=e^{i\pi \tilde{c}_{ij}}$, with
\begin{align}
\tilde{c} = 
\bordermatrix{~&S&b_1&b_2&b_3&b_4&b_5&\zeta&\alpha\cr
S&0&0&0&0&0&0&1&1\cr
b_1&1&1&1&1&1&1&1&-1/2\cr
b_2&1&1&1&1&1&1&1&-1/2\cr
b_3&1&1&1&1&1&1&1&1\cr
b_4&1&1&1&1&1&1&1&-1/2\cr
b_5&1&1&1&1&1&1&1&+1/2\cr
\zeta&1&1&1&1&1&1&1&1\cr
\alpha&1&1&1&1&1&0&1&-1/2\cr
}\,.
\end{align}
In the notation employed in \eqref{rfbasis} included fermions are periodic unless indicated otherwise (for example, $\frac{1}{2}$ denotes fermions twisted by $-i$).
The string model under consideration possesses $N=1$ space-time supersymmetry and $SU(5)\times{U(1)}\times{U(1)}^4\times{SU(4)}\times{SO(10)}$ gauge symmetry.
We use the terms ``observable" and ``hidden" gauge symmetry to refer to the $SU(5)\times{U(1)}$ and ${SU(4)}\times{SO(10)}$ group factors respectively.
The four extra abelian group factors ${U(1)}^4=\prod_{i=1}^4{U(1)}_i$ associated  with the world-sheet currents
$\oeta^1\oeta^{1\ast},\oeta^2\oeta^{1\ast},\oeta^3\oeta^{3\ast}, \overline{\omega}^2\overline{\omega}^3$ exhibit anomalies. As a matter of fact  ${\rm Tr}{U(1)}_1=-36, {\rm Tr}{U(1)}_2=-12, {\rm Tr}{U(1)}_3=-24, {\rm Tr}{U(1)}_4=-12$. However, redefining appropriately we obtain three abelian combinations free of gauge and mixed gravitational anomalies
\begin{align}
{U(1)}'_1 &= {U(1)}_3+2U(1)_4\,,\\
{U(1)}'_2 &= {U(1)}_1-3U(1)_2\,,\\
{U(1)}'_3 &= 3{U(1)}_1+{U(1)}_2+4{U(1)}_3-2{U(1)}_4\,,
\end{align}
and one anomalous $U(1)$ group factor
\begin{align}
{U(1)}_A = -3 {U(1)}_1 - {U(1)}_2+2{U(1)}_3-{U(1)}_4\ \,,\ {\rm Tr}{U(1)}_A = 180\,.
\end{align}

The massless matter spectrum consists of:
(i) ``Observable" sector matter particles listed in Table \ref{tobs}. These are 
states charged exclusively under the $SU(5)\times{U(1)}$ and ${U(1)}^4$ gauge 
group factor.  These comprise three chiral fermion families as well as a 
family/anti-family pair, residing in $SO(10)$ spinorials/anti-spinorials, 
coming from the sectors $b_1,b_2,b_3,b_4,b_5$, and three fiveplet/anti-fiveplet 
pairs, accommodated in $SO(10)$ vectorials, from the sectors $S, S+b_4+b_5$. 
Bosonic partners come from the sectors $S+b_i, i=1,\dots,4$ and $0, b_4+b_5$ 
respectively.
(ii) ``Hidden" sector matter particles, that is, states exclusively charged under ${SU(4)}\times{SO(10)}\times{U(1)}^4$ listed in Table \ref{this}. These arise from the sectors
\footnote{Here we employ a compact notation to denote several sectors contributing to the same field multiplet, for example  $(S)+ b_i +2\alpha(+\zeta)$ stands for four sectors: $b_i+2\alpha, b_i+2\alpha+\zeta, S+b_i+2\alpha,  S+b_i+2\alpha+\zeta$.}
 $(S)+b_i+2\alpha (+\zeta), i=1,\dots,4$.
(iii) Exotic fractionally charged states coming from the sectors $(S)+b_1\pm\alpha (+\zeta)$, $(S)+b_1+b_4+b_5\pm\alpha (+\zeta)$,
$(S)+b_2+b_3+b_5\pm\alpha (+\zeta)$, $(S)+b_1+b_2+b_4\pm\alpha (+\zeta)$. $(S)+b_2+b_4\pm\alpha (+\zeta)$, $(S)+b_4\pm\alpha (+\zeta)$. These are listed in Table \ref{texo}.

\begin{table}[!ht]
\centering
{\footnotesize
\begin{tabular}{|l|c|c||c|c|c|c||c|c|}
\hline
&$SU(5)$&$U(1)$&${U(1)}_1$&${U(1)}_2$&${U(1)}_3$&${U(1)}_4$&$SU(4)$&$SO(10)$\\
\hline
\multicolumn{9}{|l|}{$S$}\\
\hline
$h_1$&
${\mathbf{5}}$&$-1$&$+1$&$0$&$0$&$0$&${\mathbf{1}}$&$\mathbf{1}$\\
\hline
$\bar{h}_1$&${\overline{\mathbf{5}}}$&$+1$&$-1$&$0$&$0$&$0$&${\mathbf{1}}$&$\mathbf{1}$\\
\hline
$h_2$&
${\mathbf{5}}$&$-1$&$0$&$+1$&$0$&$0$&${\mathbf{1}}$&$\mathbf{1}$\\
\hline
$\bar{h}_2$&
${\overline{\mathbf{5}}}$&$+1$&$0$&$-1$&$0$&$0$&${\mathbf{1}}$&$\mathbf{1}$\\
\hline
$h_3$&
${\mathbf{5}}$&$-1$&$0$&$0$&$+1$&$0$&${\mathbf{1}}$&$\mathbf{1}$\\
\hline
$\bar{h}_3$&
${\overline{\mathbf{5}}}$&$+1$&$0$&$0$&$-1$&$0$&${\mathbf{1}}$&$\mathbf{1}$\\
\hline
$\Phi_{12}$&
${\mathbf{1}}$&$0$&$-1$&$+1$&$0$&$0$&${\mathbf{1}}$&$\mathbf{1}$\\
\hline
$\overline{\Phi}_{12}$&
${\mathbf{1}}$&$0$&$+1$&$-1$&$0$&$0$&${\mathbf{1}}$&$\mathbf{1}$\\
\hline
${\Phi}_{31}$&
${\mathbf{1}}$&$0$&$+1$&$0$&$-1$&$0$&${\mathbf{1}}$&$\mathbf{1}$\\
\hline
$\overline{\Phi}_{31}$&
${\mathbf{1}}$&$0$&$-1$&$0$&$+1$&$0$&${\mathbf{1}}$&$\mathbf{1}$\\
\hline
$\Phi_{23}$&
${\mathbf{1}}$&$0$&$0$&$-1$&$+1$&$0$&${\mathbf{1}}$&$\mathbf{1}$\\
\hline
$\overline{\Phi}_{23}$&
${\mathbf{1}}$&$0$&$0$&$+1$&$-1$&$0$&${\mathbf{1}}$&$\mathbf{1}$\\
\hline
$\Phi_I\,,I=1,\dots,5$&
${\mathbf{1}}$&$0$&$0$&$0$&$0$&$0$&${\mathbf{1}}$&$\mathbf{1}$\\
\hline
%$\Phi_2$&
%${\mathbf{1}}$&$0$&$0$&$0$&$0$&$0$&${\mathbf{1}}$&$\mathbf{1}$\\
%\hline
%$\Phi_3$&
%${\mathbf{1}}$&$0$&$0$&$0$&$0$&$0$&${\mathbf{1}}$&$\mathbf{1}$\\
%\hline
%$\Phi_4$&
%${\mathbf{1}}$&$0$&$0$&$0$&$0$&$0$&${\mathbf{1}}$&$\mathbf{1}$\\
%\hline
%$\Phi_5$&
%${\mathbf{1}}$&$0$&$0$&$0$&$0$&$0$&${\mathbf{1}}$&$\mathbf{1}$\\
%\hline
\multicolumn{9}{|l|}{$b_1$}\\
\hline
$F_1$&
${\mathbf{10}}$&$+\mfrac{1}{2}$&$-\mfrac{1}{2}$&$0$&$0$&$0$&${\mathbf{1}}$&$\mathbf{1}$\\
\hline
$\overline{f}_1$&
${\overline{\mathbf{5}}}$&$-\mfrac{3}{2}$&$-\mfrac{1}{2}$&$0$&$0$&$0$&${\mathbf{1}}$&$\mathbf{1}$\\
\hline
${\ell}^c_1$&
${\mathbf{1}}$&$+\mfrac{5}{2}$&$-\mfrac{1}{2}$&$0$&$0$&$0$&${\mathbf{1}}$&$\mathbf{1}$\\
\hline
\multicolumn{9}{|l|}{$b_2$}\\
\hline
$F_2$&
${\mathbf{10}}$&$+\mfrac{1}{2}$&$0$&$-\mfrac{1}{2}$&$0$&$0$&${\mathbf{1}}$&$\mathbf{1}$\\
\hline
$\overline{f}_2$&
${\overline{\mathbf{5}}}$&$-\mfrac{3}{2}$&$0$&$-\mfrac{1}{2}$&$0$&$0$&${\mathbf{1}}$&$\mathbf{1}$\\
\hline
${\ell}^c_2$&
${\mathbf{1}}$&$+\mfrac{5}{2}$&$0$&$-\mfrac{1}{2}$&$0$&$0$&${\mathbf{1}}$&$\mathbf{1}$\\
\hline
\multicolumn{9}{|l|}{$b_3$}\\
\hline
$F_3$&
${\mathbf{10}}$&$+\mfrac{1}{2}$&$0$&$0$&$+\mfrac{1}{2}$&$-\mfrac{1}{2}$&${\mathbf{1}}$&$\mathbf{1}$\\
\hline
$\overline{f}_3$&
${\overline{\mathbf{5}}}$&$-\mfrac{3}{2}$&$0$&$0$&$+\mfrac{1}{2}$&$+\mfrac{1}{2}$&${\mathbf{1}}$&$\mathbf{1}$\\
\hline
${\ell}^c_3$&
${\mathbf{1}}$&$+\mfrac{5}{2}$&$0$&$0$&$+\mfrac{1}{2}$&$+\mfrac{1}{2}$&${\mathbf{1}}$&$\mathbf{1}$\\
\hline
\multicolumn{9}{|l|}{$b_4$}\\
\hline
$F_4$&
${\mathbf{10}}$&$+\mfrac{1}{2}$&$-\mfrac{1}{2}$&$0$&$0$&$0$&${\mathbf{1}}$&$\mathbf{1}$\\
\hline
${f}_4$&
${{\mathbf{5}}}$&$+\mfrac{3}{2}$&$+\mfrac{1}{2}$&$0$&$0$&$0$&${\mathbf{1}}$&$\mathbf{1}$\\
\hline
$\overline{\ell}^c_4$&
${\mathbf{1}}$&$-\mfrac{5}{2}$&$+\mfrac{1}{2}$&$0$&$0$&$0$&${\mathbf{1}}$&$\mathbf{1}$\\
\hline
\multicolumn{9}{|l|}{$b_5$}\\
\hline
$\oF_5$&
$\overline{\mathbf{10}}$&$-\mfrac{1}{2}$&$0$&$+\mfrac{1}{2}$&$0$&$0$&${\mathbf{1}}$&$\mathbf{1}$\\
\hline
$\overline{f}_5$&
${{\mathbf{5}}}$&$-\mfrac{3}{2}$&$0$&$-\mfrac{1}{2}$&$0$&$0$&${\mathbf{1}}$&$\mathbf{1}$\\
\hline
${\ell}^c_5$&
${\mathbf{1}}$&$+\mfrac{5}{2}$&$0$&$-\mfrac{1}{2}$&$0$&$0$&${\mathbf{1}}$&$\mathbf{1}$\\
\hline
\multicolumn{9}{|l|}{$S+b_4+b_5$}\\
\hline
$h_{45}$&
${\mathbf{5}}$&$-1$&$-\mfrac{1}{2}$&$-\mfrac{1}{2}$&$0$&$0$&${\mathbf{1}}$&$\mathbf{1}$\\
\hline
$\overline{h}_{45}$&
$\overline{\mathbf{5}}$&$+1$&$+\mfrac{1}{2}$&$+\mfrac{1}{2}$&$0$&$0$&${\mathbf{1}}$&$\mathbf{1}$\\
\hline
${\phi}_{45}$&${{\mathbf{1}}}$&$0$&$+\mfrac{1}{2}$&$+\mfrac{1}{2}$&$+1$&$0$&${\mathbf{1}}$&$\mathbf{1}$\\
\hline
$\overline{\phi}_{45}$&${{\mathbf{1}}}$&$0$&$-\mfrac{1}{2}$&$-\mfrac{1}{2}$&$-1$&$0$&${\mathbf{1}}$&$\mathbf{1}$\\
\hline
${\phi}_{+}$&${{\mathbf{1}}}$&$0$&$+\mfrac{1}{2}$&$-\mfrac{1}{2}$&$0$&$+1$&${\mathbf{1}}$&$\mathbf{1}$\\
\hline
$\overline{\phi}_{+}$&${{\mathbf{1}}}$&$0$&$-\mfrac{1}{2}$&$+\mfrac{1}{2}$&$0$&$-1$&${\mathbf{1}}$&$\mathbf{1}$\\
\hline
${\phi}_{-}$&${{\mathbf{1}}}$&$0$&$+\mfrac{1}{2}$&$-\mfrac{1}{2}$&$0$&$-1$&${\mathbf{1}}$&$\mathbf{1}$\\
\hline
$\overline{\phi}_{-}$&${{\mathbf{1}}}$&$0$&$-\mfrac{1}{2}$&$+\mfrac{1}{2}$&$0$&$+1$&${\mathbf{1}}$&$\mathbf{1}$\\
\hline
${\phi}_{i}\,,i=1,\dots,4$&${{\mathbf{1}}}$&$0$&$+\mfrac{1}{2}$&$-\mfrac{1}{2}$&$0$&$0$&${\mathbf{1}}$&$\mathbf{1}$\\
\hline
$\overline{\phi}_{i}\,, i=1,\dots,4$&${{\mathbf{1}}}$&$0$&$-\mfrac{1}{2}$&$+\mfrac{1}{2}$&$0$&$0$&${\mathbf{1}}$&$\mathbf{1}$\\
\hline
%${\phi}_{2}$&${{\mathbf{1}}}$&$0$&$+\mfrac{1}{2}$&$-\mfrac{1}{2}$&$0$&$0$&${\mathbf{1}}$&$\mathbf{1}$\\
%\hline
%$\overline{\phi}_{2}$&${{\mathbf{1}}}$&$0$&$-\mfrac{1}{2}$&$+\mfrac{1}{2}$&$0$&$0$&${\mathbf{1}}$&$\mathbf{1}$\\
%\hline
%${\phi}_{3}$&${{\mathbf{1}}}$&$0$&$+\mfrac{1}{2}$&$-\mfrac{1}{2}$&$0$&$0$&${\mathbf{1}}$&$\mathbf{1}$\\
%\hline
%$\overline{\phi}_{3}$&${{\mathbf{1}}}$&$0$&$-\mfrac{1}{2}$&$+\mfrac{1}{2}$&$0$&$0$&${\mathbf{1}}$&$\mathbf{1}$\\
%\hline
%${\phi}_{4}$&${{\mathbf{1}}}$&$0$&$+\mfrac{1}{2}$&$-\mfrac{1}{2}$&$0$&$0$&${\mathbf{1}}$&$\mathbf{1}$\\
%\hline
%$\overline{\phi}_{4}$&${{\mathbf{1}}}$&$0$&$-\mfrac{1}{2}$&$+\mfrac{1}{2}$&$0$&$0$&${\mathbf{1}}$&$\mathbf{1}$\\
%\hline
\end{tabular}
}
\caption{``Observable" sector massless matter states and their $SU(5)\times{U(1)}\times{U(1)}^4\times{SU(4)}\times{SO(10)}$ quantum numbers.}
\label{tobs}
\end{table}

\begin{table}[!ht]
\centering
{\footnotesize
\begin{tabular}{|l|c|c||c|c|c|c||c|c|}
\hline
&$SU(5)$&$U(1)$&${U(1)}_1$&${U(1)}_2$&${U(1)}_3$&${U(1)}_4$&$SU(4)$&$SO(10)$\\
\hline
\multicolumn{9}{|l|}{$b_1+2\alpha\,\left(+\zeta\right)$}\\
\hline
$D_1$&
${\mathbf{1}}$&$0$&$0$&$-\mfrac{1}{2}$&$+\mfrac{1}{2}$&$0$&${\mathbf{6}}$&$\mathbf{1}$\\
\hline
$T_1$&
${\mathbf{1}}$&$0$&$0$&$-\mfrac{1}{2}$&$+\mfrac{1}{2}$&$0$&${\mathbf{1}}$&$\mathbf{10}$\\
\hline
\multicolumn{9}{|l|}{$b_2+2\alpha\,\left(+\zeta\right)$}\\
\hline
$D_2$&
${\mathbf{1}}$&$0$&$-\mfrac{1}{2}$&$0$&$+\mfrac{1}{2}$&$0$&${\mathbf{6}}$&$\mathbf{1}$\\
\hline
$T_2$&
${\mathbf{1}}$&$0$&$-\mfrac{1}{2}$&$0$&$+\mfrac{1}{2}$&$0$&${\mathbf{1}}$&$\mathbf{10}$\\
\hline
\multicolumn{9}{|l|}{$b_3+2\alpha\,\left(+\zeta\right)$}\\
\hline
$D_3$&
${\mathbf{1}}$&$0$&$-\mfrac{1}{2}$&$-\mfrac{1}{2}$&$0$&$+\mfrac{1}{2}$&${\mathbf{6}}$&$\mathbf{1}$\\
\hline
$T_3$&
${\mathbf{1}}$&$0$&$-\mfrac{1}{2}$&$-\mfrac{1}{2}$&$0$&$-\mfrac{1}{2}$&${\mathbf{1}}$&$\mathbf{10}$\\
\hline
\multicolumn{9}{|l|}{$b_4+2\alpha\,\left(+\zeta\right)$}\\
\hline
$D_4$&
${\mathbf{1}}$&$0$&$0$&$-\mfrac{1}{2}$&$+\mfrac{1}{2}$&$0$&${\mathbf{6}}$&$\mathbf{1}$\\
\hline
$T_4$&
${\mathbf{1}}$&$0$&$0$&$+\mfrac{1}{2}$&$-\mfrac{1}{2}$&$0$&${\mathbf{1}}$&$\mathbf{10}$\\
\hline
\multicolumn{9}{|l|}{$b_5+2\alpha\,\left(+\zeta\right)$}\\
\hline
$D_5$&
${\mathbf{1}}$&$0$&$+\mfrac{1}{2}$&$0$&$-\mfrac{1}{2}$&$0$&${\mathbf{6}}$&$\mathbf{1}$\\
\hline
$T_5$&
${\mathbf{1}}$&$0$&$-\mfrac{1}{2}$&$0$&$+\mfrac{1}{2}$&$0$&${\mathbf{1}}$&$\mathbf{10}$\\
\hline
\end{tabular}
}
\caption{``Hidden sector" massless matter states and their $SU(5)\times{U(1)}\times{U(1)}^4\times{SU(4)}\times{SO(10)}$ quantum numbers.}
\label{this}
\end{table}

\begin{table}[!ht]
\centering
{\footnotesize
\begin{tabular}{|l|c|c||c|c|c|c||c|c|}
\hline
&$SU(5)$&$U(1)$&${U(1)}_1$&${U(1)}_2$&${U(1)}_3$&${U(1)}_4$&$SU(4)$&$SO(10)$\\
\hline
\multicolumn{9}{|l|}{$b_1\pm\alpha\,\left(+\zeta\right)$}\\
\hline
$\overline{X}_1$&
${\mathbf{1}}$&$-\mfrac{5}{4}$&$-\mfrac{1}{4}$&$+\mfrac{1}{4}$&$+\mfrac{1}{4}$&$+\mfrac{1}{2}$&$\overline{\mathbf{4}}$&$\mathbf{1}$\\
\hline
$\overline{X}_2$&
${\mathbf{1}}$&$-\mfrac{5}{4}$&$-\mfrac{1}{4}$&$+\mfrac{1}{4}$&$+\mfrac{1}{4}$&$-\mfrac{1}{2}$&$\overline{\mathbf{4}}$&$\mathbf{1}$\\
\hline
\multicolumn{9}{|l|}{$b_1+b_4+b_5\pm\alpha\,\left(+\zeta\right)$}\\
\hline
${Y}_1$&
${\mathbf{1}}$&$+\mfrac{5}{4}$&$-\mfrac{1}{4}$&$+\mfrac{1}{4}$&$-\mfrac{1}{4}$&$+\mfrac{1}{2}$&${\mathbf{4}}$&$\mathbf{1}$\\
\hline
${Y}_2$&
${\mathbf{1}}$&$+\mfrac{5}{4}$&$-\mfrac{1}{4}$&$+\mfrac{1}{4}$&$-\mfrac{1}{4}$&$-\mfrac{1}{2}$&${\mathbf{4}}$&$\mathbf{1}$\\
\hline
\multicolumn{9}{|l|}{$b_2+b_3+b_5\pm\alpha\,\left(+\zeta\right)$}\\
\hline
%${Q}_1$&
${Z}_2$&
${\mathbf{1}}$&$+\mfrac{5}{4}$&$-\mfrac{1}{4}$&$+\mfrac{3}{4}$&$+\mfrac{1}{4}$&$0$&${\mathbf{4}}$&$\mathbf{1}$\\
\hline
%$\overline{Q}_1$&
$\overline{Z}_2$&
${\mathbf{1}}$&$-\mfrac{5}{4}$&$-\mfrac{3}{4}$&$+\mfrac{1}{4}$&$-\mfrac{1}{4}$&$0$&$\overline{\mathbf{4}}$&$\mathbf{1}$\\
\hline
\multicolumn{9}{|l|}{$b_1+b_2+b_4\pm\alpha\,\left(+\zeta\right)$}\\
\hline
${Y}_2'$&
${\mathbf{1}}$&$+\mfrac{5}{4}$&$-\mfrac{1}{4}$&$+\mfrac{1}{4}$&$-\mfrac{1}{4}$&$-\mfrac{1}{2}$&${\mathbf{4}}$&$\mathbf{1}$\\
\hline
$\overline{Y}_1$&
${\mathbf{1}}$&$-\mfrac{5}{4}$&$+\mfrac{1}{4}$&$-\mfrac{1}{4}$&$+\mfrac{1}{4}$&$-\mfrac{1}{2}$&$\overline{\mathbf{4}}$&$\mathbf{1}$\\
\hline
\multicolumn{9}{|l|}{$S+b_2+b_4\pm\alpha\,\left(+\zeta\right)$}\\
\hline
${Z}_1$&
${\mathbf{1}}$&$-\mfrac{5}{4}$&$+\mfrac{1}{4}$&$+\mfrac{1}{4}$&$-\mfrac{1}{4}$&$+\mfrac{1}{2}$&${\mathbf{4}}$&$\mathbf{1}$\\
\hline
$\overline{Z}_1$&
${\mathbf{1}}$&$+\mfrac{5}{4}$&$-\mfrac{1}{4}$&$-\mfrac{1}{4}$&$+\mfrac{1}{4}$&$-\mfrac{1}{2}$&$\overline{\mathbf{4}}$&$\mathbf{1}$\\
\hline
\multicolumn{9}{|l|}{$b_4\pm\alpha\,\left(+\zeta\right)$}\\
\hline
${X}_1$&
${\mathbf{1}}$&$+\mfrac{5}{4}$&$+\mfrac{1}{4}$&$-\mfrac{1}{4}$&$-\mfrac{1}{4}$&$-\mfrac{1}{2}$&${\mathbf{4}}$&$\mathbf{1}$\\
\hline
$\overline{X}_2'$&
${\mathbf{1}}$&$-\mfrac{5}{4}$&$-\mfrac{1}{4}$&$+\mfrac{1}{4}$&$+\mfrac{1}{4}$&$-\mfrac{1}{2}$&$\overline{\mathbf{4}}$&$\mathbf{1}$\\
\hline
\end{tabular}
}
\caption{Exotic fractionally charged massless matter states and their $SU(5)\times{U(1)}\times{U(1)}^4\times{SU(4)}\times{SO(10)}$ quantum numbers.}
\label{texo}
\end{table}

\clearpage
%%%%%%%%%%%%%%%%%%%%%%%%%%%%%%

\section{F-flatness condition analysis}
\label{appb}
In this appendix we consider solutions of the F-flatness equations of a superpotential of the form
\begin{align}\label{nrw}
w&=\oPhi_{23}\oPhi_{31}\oPhi_{12} +
\oPhi_{12} \left(\ophi_1^2+\ophi_2^2+\ophi_4^2+\ophi_+\ophi_-\right)
\nonumber\\
&+\left(D_1^2+D_4^2+T_1^2\right)\oPhi_{23}+\left(\alpha_1 \ophi_1^2+\alpha_2 \ophi_2^2 + \alpha_3 \ophi_4^2+\alpha_4\ophi_+ \ophi_-\right) D_1^2 \Phi_{31} \\
&+\left(\!\beta_1 \ophi_1^2\!+\!\beta_2 \ophi_2^2 \!+\! \beta_3\ophi_4^2\!+\!\beta_4 \ophi_+\ophi_-\!\right)\! D_4^2 \Phi_{31} \!+\! \left(\gamma_1 \ophi_1^2+\gamma_2 \ophi_2^2 + \gamma_3\ophi_4^2+\gamma_4 \ophi_+\ophi_-\right)\! T_1^2 \Phi_{31},\nonumber
%\label{nrw}
\end{align}
where $\alpha_i, \beta_i, \gamma_i, i=1,2,3,4$ are the coupling constants of the associated fifth order terms.
We focus on perturbative solutions assuming $\Phi_{31},\oPhi_{31} \sim \xi$, where $\xi$ is a small parameter in
appropriate mass units.
The F-flatness conditions read
\begin{align}
\frac{\partial w}{\partial \oPhi_{23}}&= D_1^2+D_4^2+T_1^2+\oPhi_{31}\oPhi_{12} = 0\,,\label{naoPhi23}\\
\frac{\partial w}{\partial \oPhi_{31}}&= \oPhi_{23} \oPhi_{12} = 0\,,\label{naoPhi31}\\
\frac{\partial w}{\partial \oPhi_{12}}&= \ophi_1^2+\ophi_2^2+\ophi_4^2+\ophi_+\ophi_-+\oPhi_{23}\oPhi_{31} = 0\,,
\label{naoPhi12}\\
\frac{\partial w}{\partial \Phi_{31}}&=
\left(\alpha_1 \ophi_1^2\!+\!\alpha_2\ophi_2^2\!+\!\alpha_3 \ophi_4^2\!+\!\alpha_4\ophi_+ \ophi_-\right)\! D_1^2 
\!+\!\left(\beta_1 \ophi_1^2\!+\!\beta_2\ophi_2^2\!+\!\beta_3\ophi_4^2\!+\!\beta_4 \ophi_+\ophi_-\right)\! D_4^2\nonumber\\
&+\left(\gamma_1 \ophi_1^2\!+\!\gamma_2\ophi_2^2\!+\!\gamma_3\ophi_4^2\!+\!\gamma_4 \ophi_+\ophi_-\right) T_1^2=0\,,
\label{naPhi31}\\
\frac{\partial w}{\partial \ophi_1}&=2\ophi_1 \left[\oPhi_{12}+ \Phi_{31}\left(\alpha_1 D_1^2+\beta_1 D_4^2+\gamma_1 T_1^2\right)\right]=0\,,\label{nfop1}\\
\frac{\partial w}{\partial \ophi_2}&=2\ophi_2 \left[\oPhi_{12}+ \Phi_{31}\left(\alpha_2 D_1^2+\beta_2 D_4^2+\gamma_2 T_1^2\right)\right]=0\,,\label{nfop2}\\
\frac{\partial w}{\partial \ophi_4}&=2\ophi_4 \left[\oPhi_{12}+ \Phi_{31}\left(\alpha_3 D_1^2+\beta_3 D_4^2+\gamma_3 T_1^2\right)\right]=0\,,\label{nfop4}\\
\frac{\partial w}{\partial \ophi_+}&=\ophi_- \left[\oPhi_{12}+ \Phi_{31}\left(\alpha_4 D_1^2+\beta_4 D_4^2+\gamma_4 T_1^2\right)\right]=0\,,\label{nfopp}\\
\frac{\partial w}{\partial \ophi_-}&=\ophi_+ \left[\oPhi_{12}+ \Phi_{31}\left(\alpha_4 D_1^2+\beta_4 D_4^2+\gamma_4 T_1^2\right)\right]=0\,,\label{nfopm}\\
\frac{\partial w}{\partial D_1}&=2 D_1\left[\oPhi_{23}+ \Phi_{31}\left(\alpha_1\ophi_1^2+
\alpha_2\ophi_2^2+\alpha_3\ophi_4^2+\alpha_4\ophi_+ \ophi_-\right)\right]=0\,,\label{nad1}\\
\frac{\partial w}{\partial D_4}&=2D_4\left[\oPhi_{23}+ \Phi_{31}\left(\beta_1\ophi_1^2+\beta_2\ophi_2^2+\beta_3\ophi_4^2+\beta_4\ophi_+ \ophi_-\right)\right]=0\,,\label{nad4}\\
\frac{\partial w}{\partial T_1}&=2T_1\left[\oPhi_{23}+ \Phi_{31}\left(\gamma_1\ophi_1^2+\gamma_2\ophi_2^2+\gamma_3\ophi_4^2+\gamma_4\ophi_+ \ophi_-\right)\right]=0\,.\label{nat1}
\end{align}
Using the coupling relations
\begin{align}
-\frac{\alpha_1}{3}=-\frac{\gamma_1}{3}=-\frac{\beta_1}{2}=\frac{\beta_3}{2}=
\frac{\beta_4}{2}=\alpha_4\,,\\
\alpha_2=\alpha_3=\gamma_2=\gamma_3=\gamma_4=
\alpha_4\ ,\ \beta_2=0\,,
\end{align}
Eqs. \eqref{naoPhi12}, \eqref{nad1}, \eqref{nad4} and \eqref{nat1}  take the 
form
\begin{align}
\ophi_1^2 + \ophi_2^2 +\ophi_4^2+\ophi_+ \ophi_- +\oPhi_{23} \oPhi_{31} &=0 \label{eeed}\,,\\
D_1\left[\oPhi_{23}+\Phi_{31}\alpha_4\left(-3\ophi_1^2 + \ophi_2^2+\ophi_4^2 +\ophi_+ \ophi_- \right)\right] &=0\label{eeea}\,,\\
D_4\left[\oPhi_{23}+\Phi_{31}\alpha_4\left(-2\ophi_1^2 + 2\ophi_4^2+ 2 \ophi_+ \ophi_- \right)\right] &=0\label{eeeb}\,,\\
T_1\left[\oPhi_{23}+\Phi_{31}\alpha_4\left(-3\ophi_1^2 + \ophi_2^2 +\ophi_4^2+\ophi_+ \ophi_- \right)\right] &=0\,.\label{eeec}
\end{align}
Solving \eqref{eeed} with respect to $\ophi_1^2$ and substituting into \eqref{eeea} and \eqref{eeeb}, \eqref{eeec} we get
\begin{align}
D_1\left[\oPhi_{23}+\Phi_{31}\alpha_4\left(4\ophi_2^2+4\ophi_4^2 + 4\ophi_+ \ophi_- \right)\right] &=0\label{eeean}\,,\\
D_4\left[\oPhi_{23}+\Phi_{31}\alpha_4\left(2\ophi_2^2+4\ophi_4^2+ 4 \ophi_+ 4\ophi_- \right)\right] &=0\label{eeebn}
\,,
\\
T_1\left[\oPhi_{23}+\Phi_{31}\alpha_4\left(4\ophi_2^2+4\ophi_4^2+ 4 \ophi_+ 4\ophi_- \right)\right] &=0\label{eeecn}
\,,
\end{align}
where we have neglected terms of order $\xi^5$.

Moreover, Eqs. \eqref{naoPhi23} and \eqref{nfop1}, \eqref{nfop2}, \eqref{nfop4}, \eqref{nfopp}  yield
\begin{align}
D_1^2+D_4^2+T_1^2+\oPhi_{31}\oPhi_{12}&=0\label{DTD}\,,\\
\ophi_1\left[\oPhi_{12} - \alpha_4 \Phi_{31}\left(3 D_1^2 +2 D_4^2+3 T_1^2 
\right)\right]&=0\label{DTA}\,,\\
\ophi_2\left[\oPhi_{12} + \alpha_4 \Phi_{31}\left(\oPhi_{12}+ D_1^2 + T_1^2 \right)\right]&=0\label{DTB}\,,\\
\ophi_4\left[\oPhi_{12} + \alpha_4 \Phi_{31}\left(\oPhi_{12}+ D_1^2 + 
2D_4^2+T_1^2 \right)\right]&=0\label{DTC}\,,\\
\ophi_+\left[\oPhi_{12} + \alpha_4 \Phi_{31}\left(\oPhi_{12}+ D_1^2 + 
2D_4^2+T_1^2 \right)\right]&=0\label{DTE}\,.
\end{align}
Solving \eqref{DTD} with respect to $T_1^2$ and substituting into \eqref{DTA}-\eqref{DTE} we get
\begin{align}
\ophi_1\left[\oPhi_{12} + \alpha_4 \Phi_{31} D_4^2\right]&=0\label{DTAn}\,,\\
\ophi_2\left[\oPhi_{12} - \alpha_4 \Phi_{31} D_4^2\right]&=0\label{DTBn}\,,\\
\ophi_4\left[\oPhi_{12} + \alpha_4 \Phi_{31} D_4^2\right]&=0\label{DTAnm}\,,\\
\ophi_+\left[\oPhi_{12} + \alpha_4 \Phi_{31} D_4^2\right]&=0\label{DTAnmm}\,.
\end{align}
These are not compatible unless $\ophi_2\lesssim \xi^2$. In this case
\begin{align}
\oPhi_{12} &= - \alpha_4 D_4^2 \Phi_{31}\,,\\
T_1^2&=-D_1^2-D_4^2 - \oPhi_{31}\oPhi_{12} \,,
\end{align}
and the system \eqref{eeean}-\eqref{eeecn} also yields
\begin{align}
\oPhi_{23}&=-2\Phi_{31}\alpha_4\left(-\ophi_1^2 + \ophi_4^2+  \ophi_+ \ophi_- \right)\,,\\
\ophi_1^2 &=  -\ophi_4^2-\ophi_+ \ophi_- -\ophi_2^2-\oPhi_{23} \oPhi_{31}\,.  
\end{align}
Finally, we consider \eqref{naPhi31}. Using, \eqref{nad1}, \eqref{nad4}, \eqref{nat1} all three factors in parentheses equal to 
$-\frac{\oPhi_{23}}{\Phi_{31}}$. That is
\begin{align}
-\frac{\oPhi_{23}}{\Phi_{31}}\left(D_1^2+D_4^2+T_1^2\right) =0 \Rightarrow
\frac{\oPhi_{23}\oPhi_{12} \oPhi_{31}}{\Phi_{31}} = 0
\end{align}
which is satisfied at order $\xi^6$ by the use of \eqref{naoPhi23}. The same holds for \eqref{naoPhi31}.
\newpage

\section{ Operator product expansions, correlators etc}
\label{appcoup}
The conformal correlator functions appearing in Section 5 can be calculated 
using techniques explained in \cite{Kalara:1990fb},\cite{DiFrancesco:1987ez}. 
These include correlators 
of exponentials which can be easily calculated using
\begin{align}
\langle{\prod_{i}e^{i \alpha_i \phi(z_i)}}\rangle=\prod_{i<j} (z_{ij})^{\alpha_i \alpha_j}\,,
\end{align}
the ghost correlator given by
\begin{align}
\langle{e^{-c/2}(z_1)}{e^{-c/2}(z_2)}{e^{-c}(z_3)}\rangle=z_{12}^{-1/4} 
z_{13}^{-1/2} z_{23}^{-1}\,,
\end{align}
and the space-time spinor correlator given by
\begin{align}
\langle{S_a(z_1) S_b(z_2)}\rangle = C_{ab} z_{12}^{-1/2}\,.
\end{align}
In addition, we use the following Ising type correlators 
\begin{align}
\langle{\sigma_+(1)\sigma_+(2)}\rangle &= 
\langle{\sigma_-(1)\sigma_-(2)}\rangle = \left|z_{12}\right|^{-1/4}\,,\\
\langle{f(1)f(2)}\rangle &= \langle{\bar{f}(1)\bar{f}(2)}\rangle = 
\frac{1}{z_{12}}\,,\\ 
\langle{\sigma_+(1)\sigma_-(2) f(3)}\rangle &= 
\frac{e^{i\pi/4}}{\sqrt{2}} z_{12}^{+3/8} z_{13}^{-1/2} z_{23}^{-1/2} \bar{z}_{12}^{-1/8} 
\\\langle{\sigma_+(1)\sigma_-(2) \bar{f}(3)}\rangle &= 
\frac{e^{-i\pi/4}}{\sqrt{2}} \bar{z}_{12}^{+3/8} \bar{z}_{13}^{-1/2} 
\bar{z}_{23}^{-1/2} {z}_{12}^{-1/8}\,,
\\
\langle{\sigma_+(1)\sigma_+(2)\sigma_+(3)\sigma_+(4)}\rangle &= 
\frac{1}{\sqrt{2}} \frac{\left|z_{13}z_{24}\right|^{1/4}}
{\left|z_{14}z_{23}z_{12}z_{34}\right|^{1/4}}\left(|x|+|1-x|+1\right)^{1/2}\\
&=\frac{1}{\sqrt{2}} \left|z_{13}z_{24}\right|^{-1/4}\left|x(1-x)\right|^{-1/4} 
\left(|x|+|1-x|+1\right)^{1/2}\,,
\\
\langle{\sigma_+(1)\sigma_+(2)\sigma_-(3)\sigma_-(4)}\rangle
&= \frac{1}{\sqrt{2}} \left|z_{12}z_{34}\right|^{-1/4} \left|1-x\right|^{-1/4} 
\left(|1-x|-|x|+1\right)^{1/2}
\,,
\\
\langle{\sigma_+(1) \sigma_+(2){f}(3){f}(4) }\rangle &=
\frac{1}{2}
\frac{\left|z_{12}\right|^{-1/4}}{{z}_{34}}\frac{2-{x}}{\sqrt{1-{x}}}
\,,
\\
\langle{\sigma_+(1) \sigma_+(2)\bar{f}(3)\bar{f}(4) }\rangle &=
\frac{1}{2}
\frac{\left|z_{12}\right|^{-1/4}}{\bar{z}_{34}}\frac{2-\bar{x}}{\sqrt{1-\bar{x}}}\,,
\end{align}
where
\begin{align}
x= \frac{z_{12} z_{34}}{z_{13}z_{24}}\,.
\end{align}

\end{appendices}

\bibliographystyle{utphys}
\providecommand{\href}[2]{#2}\begingroup\raggedright\endgroup

\end{document}